\documentclass[journal, onecolumn, 12pt, draftcls]{IEEEtran}

\usepackage{cite}
\usepackage{enumerate}
\usepackage{graphicx}
\usepackage[cmex10]{amsmath} 
\usepackage{amssymb}
\usepackage{mathtools}
\usepackage{xspace}
\usepackage{subfigure}
\usepackage[ruled,vlined]{algorithm2e} 
\usepackage{colonequals}
\usepackage{multirow}
\usepackage{siunitx}

\usepackage{epsfig}
\usepackage{epstopdf}

\usepackage{tikz}

\usepackage{pgfplots}

\usetikzlibrary{arrows,shapes,backgrounds,plotmarks,positioning}

\pgfplotsset{compat=newest}                         
\pgfplotsset{plot coordinates/math parser=false}
\newlength\figureheight
\newlength\figurewidth

\newtheorem{theorem}{Theorem}[section]

\newtheorem{definition}[theorem]{Definition}

\newtheorem{proposition}[theorem]{Proposition}
\newtheorem{assumption}[theorem]{Assumption}
\newtheorem{remark}[theorem]{Remark}

\newcommand{\Z}{\mathbb{Z}}
\newcommand{\R}{\mathbb{R}}
\newcommand{\snr}{\gamma}
\newcommand{\SNR}{\Gamma}

\newcommand{\IND}{\mathbb{I}}

\newcommand{\A}{\mathcal{A}}

\newcommand{\HSet}{\mathcal{H}}   
\newcommand{\B}{\mathcal{B}}         

\newcommand{\LN}{L^{\natural}}
\newcommand{\One}{\mathbf{1}}
\newcommand{\Zero}{\mathbf{0}}

\newcommand{\X}{\mathcal{X}}

\newcommand{\x}{\mathbf{x}}
\newcommand{\y}{\mathbf{y}}
\newcommand{\act}{\mathbf{a}}

\newcommand{\gv}{\mathbf{g}}

\newcommand{\ev}{\mathbf{e}}

\newcommand{\Th}{\phi}  
\newcommand{\Thv}{\boldsymbol{\phi}}  
\newcommand{\ThSet}{\Phi}         

\newcommand{\Charact}{\chi_{\scalebox{0.6}{$U_d$}}^{\phantom{l}}}

\newcommand{\E}{\mathbb{E}}

\begin{document}

\title{Discrete Convexity and Stochastic Approximation for Cross-layer On-off Transmission Control}

\author{Ni~Ding,~\IEEEmembership{Student Member,~IEEE}, Parastoo~Sadeghi,~\IEEEmembership{Senior Member,~IEEE}, Rodney~A.~Kennedy,~\IEEEmembership{Fellow,~IEEE}
\thanks{The authors are with the Research School of Engineering, College of Engineering and Computer Science, the Australian National University (ANU), Canberra, ACT 0200, Australia (email: $\{$ni.ding, rodney.kennedy, parastoo.sadeghi$\}$@anu.edu.au).}
}

\markboth{IEEE Transactions on Wireless Communications}%
{Ding \MakeLowercase{\textit{et al.}}: Discrete Convexity and Stochastic Approximation for Cross-layer On-off Transmission Control}

\maketitle

\begin{abstract}
This paper considers the discrete convexity of a cross-layer on-off transmission control problem in wireless communications. In this system, a scheduler decides whether or not to transmit in order to optimize the long-term quality of service (QoS) incurred by the queueing effects in the data link layer and the transmission power consumption in the physical (PHY) layer simultaneously. Using a Markov decision process (MDP) formulation, we show that the optimal policy can be determined by solving a minimization problem over a set of queue thresholds if the dynamic programming (DP) is submodular. We prove that this minimization problem is discrete convex. In order to search the minimizer, we consider two discrete stochastic approximation (DSA) algorithms: discrete simultaneous perturbation stochastic approximation (DSPSA) and $\LN$-convex stochastic approximation ($\LN$-convex SA). Through numerical studies, we show that the two DSA algorithms converge significantly faster than the existing continuous simultaneous perturbation stochastic approximation (CSPSA) algorithm in multi-user systems. Finally, we compare the convergence results and complexity of two DSA and CSPSA algorithms where we show that DSPSA achieves the best trade-off between complexity and accuracy in multi-user systems.
\end{abstract}

\begin{IEEEkeywords}
convergence, cross-layer optimization, discrete convexity, discrete stochastic approximation, dynamic programming, Markov processes.
\end{IEEEkeywords}

\ifCLASSOPTIONpeerreview
\begin{center} \bfseries EDICS Category: 3-BBND \end{center}
\fi

\IEEEpeerreviewmaketitle

\section{Introduction}

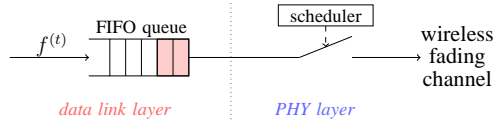
\begin{figure}[tb]
	\centering
		\centerline{\scalebox{0.7}{\begin{tikzpicture}

\draw [ ->] (0,0) -- (1.5,0);
\node at (0.8,0.25) {$f^{(t)}$};

\node at (6,0.8) {\small scheduler};
\draw (5.1,0.6) rectangle (6.9,1);
\draw [dashed,->] (6,0.6) -- (6,0.2);

\draw [fill=red!20] (2.8,0.35) rectangle (3.4,-0.35);
\draw (1.9,0.35) rectangle (3.4,-0.35);
\draw (1.9,0.35) -- (1.5,0.35);
\draw (1.9,-0.35) -- (1.5,-0.35);
\draw (2.2,0.35) -- (2.2,-0.35);
\draw (2.5,0.35) -- (2.5,-0.35);
\draw (2.8,0.35) -- (2.8,-0.35);
\draw (3.1,0.35) -- (3.1,-0.35);
\node at (2.5,0.55) {\small FIFO queue};

\draw (3.4,0) -- (5.5,0);
\draw (5.5,0) -- (6.5,0.4);

\draw [->] (6.5,0) -- (7.8,0);

\node at (8.5,0.45) {wireless};
\node at (8.5,0) {fading };
\node at (8.5,-0.4) {channel };

\draw[dotted,color=violet] (4.2,1.2)--(4.2,-1.2);
\node at (2,-1) [color=red!60] {\small \textit{data link layer}};
\node at (5.8,-1) [color=blue!60] {\small \textit{PHY layer}};

\end{tikzpicture}}}
	\caption{Cross-layer on-off transmission control in wireless communications: A scheduler decides whether or not to transmit a packet in the queue according to the optimization concerns in both layers, e.g., packet delay and queue overflow in data link layer and transmission error and spectral efficiency in physical (PHY) layer, in the long run.}
	\label{fig:QG}
\end{figure}

Consider the communication system in Fig.~\ref{fig:QG}. It is assumed that messages encapsulated in equal length packets from a higher layer (say, application layer) arrive at data link layer randomly. The packets are temporarily stored in a first-in-first-out (FIFO) queue before the transmission in the physical (PHY) layer. The departure of the queue is controlled by a scheduler: If the switch is open, no packet departs from the queue; If the switch is closed, a unit packet departs from the queue and is transmitted through the wireless channel. The objective is to find a policy or decision rule that optimizes packet delay and/or queue overflow in the data link layer and the transmission error rate and/or spectral efficiency in the PHY layer simultaneously and in the long run.

The problem in Fig.~\ref{fig:QG} is a cross-layer transmission control one because it not only incorporates the idea of rate adaptation in the PHY layer \cite{Goldsmith1997,Alouini2000} but also takes into account the quality of service (QoS) incurred by the queueing effects in the data link layer. Since the general cross-layer rate adaptation problem usually allows the scheduler to choose from a set of transmission rates \cite{Liu2005,Liu2005MultiU}, the problem in Fig.~\ref{fig:QG} can be considered as a special case where the scheduler only makes binary decisions: whether or not to transmit. We call it cross-layer on-off transmission control. This problem has been presented in some buffer scheduling problems in wireless communications, e.g., \cite{Ngo2009}, and is commonly seen in studies on network-coded relaying systems, e.g., \cite{ChenONC2007,HsuONC2011,Ding2012,ChenBANC2012}.

By assuming an \textit{i.i.d.}\ message arrival process and a finite-state Markov chain (FSMC) \cite{Sadeghi2008} modeled channel, the system in Fig.~\ref{fig:QG} is usually modeled by a Markov decision process (MDP), and dynamic programming (DP) algorithms are used to search the optimal policy, e.g., \cite{Ding2012,Ngo2009}. DP is a classic algorithm for solving MDP modeled sequential decision making problems. However, the crucial limitation of DP is that its computation load grows drastically with the cardinalities of the state sets in MDP. This problem is called the \textit{curse of dimensionality} \cite{SuttonRL1998} and makes DP inefficient for solving high dimensional MDP problems. To relieve the complexity, most related studies, e.g., \cite{Djonin2007,Djonin2007Struct,Ngo2009,Huang2010,HsuONC2011,Ding2013Mono,Ngo2010}, focus on proving the monotonicity of the optimal policy in queue occupancy/state. It is because that in this case the optimal policy is a switching curve or plane that is fully characterized by a set of optimal queue thresholds. These optimal queue thresholds can be searched by solving a multivariate minimization problem with much lower complexity than DP. To approximate the optimizer of this problem, a stochastic approximation (SA) method is usually considered. The typical examples are the continuous simultaneous perturbation stochastic (CSPSA) algorithms proposed in \cite{Huang2010,Ngo2010,Bhatnagar2011}.
But, there are two problems with these SA algorithms.
One is that the authors in \cite{Huang2010,Ngo2010,Bhatnagar2011} only apply SA without showing or analyzing the convexity of the objective function. SA is based on an iterative line search method. When it applies to a non-convex minimization problem, it may just converge to the local optimizer with probability. There are some studies showing the sufficient conditions for the global and/or almost sure convergence of SA algorithms. But, as pointed out in \cite{LimDSA2012}, these conditions are usually difficult to verify for a non-convex objective function.\footnote{Spall showed in \cite{Spall1992mult} the sufficient conditions for SPSA\cite{SpallSPSA1998} to converge almost surely in a continuous optimization problem. They require the objective function to be differentiable and the estimation sequence generated by a gradient descent method to converge to the optimizer. Most of the discrete SPSA algorithms are adapted from \cite{SpallSPSA1998}, e.g., \cite{Gerencser1999}. Usually, the convergence performance is conditioned on certain property of the subgradient and is not easy to verify, e.g., Theorem $1$ in \cite{Bhatnagar2011}.}
On the other hand, if one can prove the convexity of the objective function, these conditions are usually straightforwardly satisfied. In addition, there exists SA algorithms that are exclusively proposed for discrete convex minimization problems in the literature for which the global and almost sure convergence is guaranteed, e.g., \cite{Hill2004}.
The other problem with the SA algorithms in \cite{Huang2010,Ngo2010,Bhatnagar2011} is that CSPSA is originally proposed for continuous minimization problems. When it is applied to discrete ones, one needs to solve the problem of how to estimate the value of the objective function at real-valued points. One solution as proposed in \cite{Huang2010,Bhatnagar2011} is to use a projection function to project the real-valued points to integer ones. But, the projection function adds extra complexity when implementing the CSPSA algorithms. In addition, if the discrete convexity of the minimization problem is proved, one does not know if the projection function has an effect on the existence of discrete convexity or the accuracy of gradient estimation in SA algorithms.

The main purpose of this paper is to prove the discrete convexity of the on-off transmission control problem in Fig.~\ref{fig:QG} and show that this problem can be solved more efficiently by discrete stochastic approximation (DSA) algorithms than CSPSA. In this paper, we first follow the same approach as in \cite{Huang2010,Ngo2010}: We prove that the optimal transmission policy is monotonic in queue states and can be expressed by a queue threshold vector if DP is submodular. We formulate the optimal transmission control problem in Fig.~\ref{fig:QG} as a multivariate minimization problem over a set of queue thresholds. But, before proposing the solutions, we observe the shape of the objective function and prove that it is discrete convex. We then consider two discrete stochastic approximation (DSA) algorithms for searching the optimal policy: discrete simultaneous perturbation stochastic approximation (DSPSA)\cite{Wang2011} and $\LN$-convex SA \cite{LimDSA2012}. We run experiments on three systems to show the convergence performance of two DSA algorithms. The results are compared to a CSPSA algorithm. The main results in this paper are listed as follows:

\begin{itemize}
    \item For the transmission control problem in Fig.~\ref{fig:QG}, we derive a sufficient condition for the optimal policy to be nondecreasing in queue states: the submodularity of DP function. We show that the monotonic optimal transmission policy can be determined by a queue threshold vector. Each dimension of this vector determines the queue state when the transmission policy changes from \lq{not transmit}\rq\ to \lq{transmit}\rq\ when the channel is in a certain state.
    \item We convert DP to a stochastic minimization problem over queue threshold vectors and prove that the objective function is both discrete separable convex and $\LN$-convex.
    \item We present a DSPSA algorithm and an $\LN$-convex SA algorithm. Due to the discrete convexity of the minimization problem under consideration, both of them are able to converge almost surely to the optimal queue threshold vector. We run the two algorithms in single-user and two multi-user on-off transmission control systems. The results are compared to a CSPSA algorithm that uses the projection function proposed in \cite{Bhatnagar2011}.
    \item We also analyze the accuracy and complexity of two DSA algorithms and the CSPSA algorithm based on numerical experiment results. There is a tradeoff between accuracy and complexity: DSPSA and CSPSA requires less measurements of the objective function in each iteration but converges slower than $\LN$-convex SA; $\LN$-convex SA generates more accurate estimation sequence of the optimizer but requires more measurements of the objective function than DSPSA and CSPSA. Also, DSPSA converges faster than CSPSA in multi-user systems. These results can be used to guide the implementation of SA algorithms in real applications: If one can prove the discrete convexity of an on-off cross-layer transmission control problem, DSPSA and $\LN$-convex SA are more efficient than CSPSA; If the system is a multi-user one, DSPSA achieves the best trade-off between complexity and accuracy.
\end{itemize}

\begin{table}[tbp]
	\renewcommand{\arraystretch}{1.3}
	\caption{Notations}
	\label{tab:SymbMDP}
	\centering
	\small
	\begin{tabular}{c l}
	\hline\hline
	\bfseries{symbol} & \multicolumn{1}{c}{\bfseries{description}} \\ %
	\hline
    $t$ & decision epoch \\
	$f$ & the number of inflow packets\\
    $L$ & queue length (in packets)\\
	$\snr$ & signal-to-noise ratio (SNR)  \\
	$\SNR_k$ & the $k$th SNR boundary in FSMC \\
    $K$ & the number of channel states in FSMC \\
	$h,\HSet$ & channel state, channel state set\\
    $b$, $\B$ & queue state/occupancy, queue state set\\
    $\x$, $\X$ & system state, system state set\\
    $a$, $\A$ & action, action set\\
    $P_{hh'}$ & channel state transition probability from $h$ to $h'$ \\
    $P_{bb'}^{a}$ & queue state transition probability from $b$ to $b'$ under action $a$\\
    $P_{\x\x'}^{a}$ & system state transition probability from $\x$ to $\x'$ under action $a$\\
    $c_q(b,a)$ & immediate cost in the data link layer\\
    $c_{tr}(h,a)$ & immediate cost in the PHY layer\\
    $c(\x,a)$ & immediate cost in the entire system\\
    $\theta(\x)$ & stationary and deterministic policy\\
    $V_\theta(\x)$ & expected long-term cost under policy $\theta$ \\
    $\beta$ & discount factor \\ [0.3ex]
    $\Thv$& queue threshold vector\\
	$\Th_h$ & the $h$th tuple in $\Thv$\\
	$\tilde{\Thv}^{(n)}$ & the $n$th estimation of the minimizer\\
    $\gv(\tilde{\Thv}^{(n)})$ & descent direction at $\tilde{\Thv}^{(n)}$ \\
    $A$, $B$, $C$, $\alpha$, $\gamma$ & step size parameters \\
    $J(\Thv)$ & objective function at $\Thv$ \\
	$\hat{J}(\Thv)$ & noisy measurement of $J$  \\
    $\tilde{J}(\Thv)$ & piecewise linear interpolation (PLI) of $J$ \\ [0.3ex]
	\hline 
	\end{tabular}
\end{table}

\subsection{Organization}

The rest of the paper is organized as follows. In Section~\ref{sec:MDP}, we describe the MDP formulation, state the objective and present the DP algorithm for the system model in Fig.~\ref{fig:QG}.  In Section~\ref{sec:DPtoLconvex}, we prove the monotonicity of the optimal transmission policy and formulate a discrete convex optimization problem based on the submodularity of DP. In Section~\ref{sec:SA}, we present DSPSA and $\LN$-convex SA algorithms and describe their implementation details. In Section~\ref{sec:NumRes}, we apply DSPSA, $\LN$-convex SA and CSPSA algorithms to single-user and multi-user systems. The accuracy and complexity of these three algorithms are analyzed.

\subsection{Notation}

In this paper, we use $\R_+$ and $\Z$ to denote nonnegative real number set and integer number set, respectively. In TABLE~\ref{tab:SymbMDP}, we list the descriptions of symbolic notations that are used in Sections~\ref{sec:MDP}, \ref{sec:DPtoLconvex} and \ref{sec:SA}. In the MDP formulation in Section~\ref{sec:MDP}, we use superscript $(t)$ to denote the variable at decision epoch $t$, e.g., $\snr^{(t)}$ denotes the instantaneous SNR value at $t$. In the multi-user systems in Section~\ref{sec:NumRes}, we use the subscript $i$ to denote the variable of user $i$, e.g., $\snr_i^{(t)}$ denotes the instantaneous SNR value of the channel of user $i$ at $t$.

\section{MDP Formulation and Dynamic Programming}
\label{sec:MDP}

Consider the transmission control system with wireless multipath fading channel in Fig.~\ref{fig:QG}. Let time be divided into small intervals, called \textit{decision epochs} and denoted by $t$. The decision making process is infinitely long, i.e., $t\in\{0,1,\dotsc,\infty\}$. We assume the followings in this system.

\begin{assumption} \label{ass1}
Let $\{f^{(t)}\}$ be an \textit{i.i.d.}\ random message arrival process, where $f^{(t)}$ denotes the number of packets that arrive at the FIFO queue at $t$. The scheduler knows the statistics of $\{f^{(t)}\}$.
\end{assumption}
\begin{assumption} \label{ass2}
Denote $\snr^{(t)}$ the instantaneous signal-to-noise ratio (SNR) of the fading channel. Let $\{\snr^{(t)}\}$ be a random process that is independent of $\{f^{(t)}\}$. The full variation range of $\snr^{(t)}$ is partitioned into $K$ non-overlapping regions $\{[\SNR_1,\SNR_2),[\SNR_2,\SNR_3),\dotsc,[\SNR_{K},\SNR_{K+1})\}$, where $\SNR_{K+1}=\infty$. Region $[\SNR_k,\SNR_{k+1})$ is called channel state $k$. Denote $h^{(t)}$ as the channel state variable at decision epoch $t$. We say that $h^{(t)}=k$ if $\snr^{(t)}\in{[\SNR_{k},\SNR_{k+1})}$. Let the channel be modeled by an FSMC\cite{Sadeghi2008}, where $P_{h^{(t)}h^{(t+1)}}=\Pr(h^{(t+1)}|h^{(t)})$, the channel state transition probability, is determined by channel parameters and statistics and is stationary (time invariant). The scheduler knows the statistics of $\{h^{(t)}\}$ and has the real-time information on channel state, the value of $h^{(t)}$, to support the decisions.
\end{assumption}
\begin{assumption} \label{ass3}
Let $a^{(t)}\in{\A}=\{0,1\}$ be the action taken by the scheduler at $t$, where $0$ denotes \lq{not transmit}\rq\ and $1$ denotes \lq{transmit}\rq. Whenever $a^{(t)}=1$, one packet is sent.
\end{assumption}

\subsection{MDP Modeling}

Let the system in Fig.~\ref{fig:QG} be modeled by a discounted infinite-horizon MDP. The system state at $t$ is $\x^{(t)}=(b^{(t)},h^{(t)})\in{\X}=\B\times{\HSet}$, where $\times$ denotes the Cartesian product. Let $L$ be the queue length, the maximum number of packets that can be stored in the queue. $b^{(t)}\in{\B}=\{0,1,\dotsc,L\}$ is called the queue occupancy/state that denotes the number of packets stored in the queue at $t$. $h^{(t)}\in{\HSet}=\{1,2,\dotsc,K\}$ is the channel state as described in Assumption~\ref{ass2}. The state transition probability $P_{\x^{(t)}\x^{(t+1)}}^{a^{(t)}}=\Pr(\x^{(t+1)}|\x^{(t)},a^{(t)})$ is given by
\begin{equation}
P_{\x^{(t)}\x^{(t+1)}}^{a^{(t)}}=P_{b^{(t)}b^{(t+1)}}^{a^{(t)}}P_{h^{(t)}h^{(t+1)}}.
\end{equation}
$P_{b^{(t)}b^{(t+1)}}^{a^{(t)}}$ is the queue state transition probability that is derived as follows.

At each decision epoch $t$, the scheduler makes a decision $a^{(t)}$, and then $f^{(t)}$ packets flow into the queue. If the queue is full, the overflow packets will be dropped. Let $[x]^+=\max\{x,0\}$. The variation of queue state can be described by the Lindley equation \cite{Asmussen2003}
\begin{equation} \label{eq:Lindley}
b \colonequals \min\{[b-a]^{+}+f,L\}.
\end{equation}
The queue state transition probability $P_{b^{(t)}b^{(t+1)}}^{a^{(t)}}=\Pr(b^{(t+1)}|b^{(t)},a^{(t)})$ can be determined by the statistics of $\{f^{(t)}\}$ as
\begin{equation} \label{eq:Pbb}
    P_{b^{(t)}b^{(t+1)}}^{a^{(t)}} =\begin{cases}
                                        \Pr \Big( f^{(t)}=b^{(t+1)}-[b^{(t)}-a^{(t)}]^{+} \Big) & b^{(t+1)}<L \\
                                        \sum_{l=L-[b^{(t)}-a^{(t)}]^{+}}\Pr(f^{(t)}=l)    &  b^{(t+1)}=L
                                \end{cases}.
\end{equation}
The immediate cost $c:\X\times{\A}\mapsto{\R_+}$ is the cost incurred immediately after the action $a^{(t)}$ and is defined as
    \begin{equation} \label{eq:ImmCost}
        c(\x^{(t)},a^{(t)})=c_q(b^{(t)},a^{(t)})+c_{tr}(h^{(t)},a^{(t)}).
    \end{equation}
$c(\x,a)$ contains two parts: $c_q$ quantifies the loss in the data link layer; $c_{tr}$ quantifies the loss in the PHY layer. We define $c_q$ as
\begin{equation} \label{eq:Cb1}
    c_q(b^{(t)},a^{(t)})=w \E_f \Big[ \big[ [b^{(t)}-a^{(t)}]^{+}+f^{(t)}-L \big]^{+} \Big],
\end{equation}
where $w>0$ is a weight factor.\footnote{Weight factor $w$ can be considered as the priority of minimizing the loss incurred in the data-link layer as opposed to the loss incurred in the PHY layer.} $c_q$ is proportional to the expected number of lost packets due to the queue overflow. We define $c_{tr}$ as
\begin{equation}  \label{eq:Ctr1}
    c_{tr}(h^{(t)},a^{(t)})=\frac{a^{(t)}(\text{erfc}^{-1}(2\bar{P}_b))^2}{\SNR_{h^{(t)}}}.
\end{equation}
$c_{tr}$ is an estimation of the minimum power required to transmit a packet with binary phase-shift keying (BPSK) modulation in channel state $h$ that will result in an average bit-error-rate (BER) no greater than $\bar{P_b}$.\footnote{$c_{tr}$ is derived based on $\bar{P}_b=\frac{1}{2}\text{erfc}(\sqrt{P_{tr}\gamma})$, which determines the average BER when transmitting BPSK packets with power $P_{tr}$ through a channel whose SNR is $\gamma$.}

\subsection{Long-term Objective and Dynamic Programming}

Let $\theta\colon\X\mapsto\A$ be a stationary deterministic policy. Denote the expected total discounted cost under policy $\theta$ as
\begin{equation} \label{eq:Vtheta}
V_{\theta}(\x)=\E \bigg[ \sum_{t=0}^{\infty} \beta^t c(\x^{(t)},\theta(\x^{(t)})) \Big| \x^{(0)}=\x \bigg].
\end{equation}
Here, $\beta\in[0,1)$ is the discount factor that ensures the convergence of the infinite series. It also describes how farsighted a decision-maker is since $\beta^t$ assigns exponentially decaying weights to the costs in the future\cite{SuttonRL1998}. The objective of the transmission control problem in Fig.~\ref{fig:QG} is to minimize the long-term losses incurred in data-link and PHY layers, which can be described as
\begin{equation} \label{eq:obj}
\min_{\theta} V_{\theta}(\x),\quad \forall{\x}\in{\X}.
\end{equation}
It is shown in \cite{Dreyfus2002} that \eqref{eq:obj} can be solved by DP \cite{SuttonRL1998}
\begin{equation} \label{eq:V}
V(\x) \colonequals \min_{a} \Big\{  c(\x,a)+\beta\sum_{\x'}P_{\x\x'}^{a}V(\x')   \Big\}, \forall{\x}.
\end{equation}
Let $n$ denote the iteration index. The sequence $\{V^{(n)}(\x)\}$ generated by \eqref{eq:V} converges for all $\x$\cite{PutermanMDP1994}. Usually, a small threshold $\epsilon>0$ is applied so that iteration \eqref{eq:V} terminates if $|V^{(N-1)}(\x)-V^{(N)}(\x)|\leq{\epsilon}$ for all $\x$ with $N<\infty$. In this paper, we use $\epsilon=10^{-4}$. The optimal policy $\theta^*$ is determined by
\begin{equation}
    \theta^*(\x)=\underset{a\in\A}{\operatorname{argmin}} \Big\{  c(\x,a)+\beta\sum_{\x'}P_{\x\x'}^{a}V^{(N)}(\x')   \Big\}, \forall{\x},
\end{equation}
To assist the analysis in Section~\ref{sec:DPtoLconvex}, we define an auxiliary function $Q$ as the minimand in \eqref{eq:V}, i.e.,
\begin{equation} \label{eq:Q}
Q(\x,a) = c(\x,a)+\beta\sum_{\x'}P_{\x\x'}^{a}V(\x').
\end{equation}
Since the MDP under consideration is stationary, we drop the notation $t$ in \eqref{eq:V} to \eqref{eq:Q} and use $\x$ and $\x'$  to denote variables at the current and next decision epochs, respectively. We will do so in the rest of the paper.

Consider the DP algorithm described in \eqref{eq:V}. In each iteration, to do the minimization in \eqref{eq:V}, every combination of the state variables must be considered, which give rise to two problems. One is the curse of dimensionality \cite{SuttonRL1998}: The time complexity grows drastically with the cardinality or the dimension of the state space in MDP. The other is that the full knowledge (including the state space and the state transition probabilities) of MDP should be known before running DP, which makes DP unsuitable for online applications. In the next section and Section~\ref{sec:SA}, we show that problem \eqref{eq:obj} can be solved by DSA algorithms. The DSA algorithms involve lower complexity than DP and are suitable for online applications since they are simulation-based algorithms. We will discuss the advantages of DSA algorithms over DP in detail in Section~\ref{sec:complex}.

\section{Discrete Convex Optimization}
\label{sec:DPtoLconvex}

In this section, we show that problem \eqref{eq:obj} can be converted to a discrete convex optimization problem due to the submodularity of DP.

\subsection{Preliminaries}
We first introduce some concepts concerning the definition of discrete convexity. For a multivariate discrete function, there are different ways to define the convexity. We consider two of them: discrete separable convexity and $\LN$-convexity.
\begin{definition}[discrete separable convexity \cite{Murota2005}] \label{def:DiscSep}
Let $f(\x)=\sum_{d=1}^{D}f_d(x_d)$, where $f\colon\Z^{D}\mapsto{\R}_+$, $f_d\colon\Z\mapsto\R_+$ for all $d$ and $\x=(x_1,\dotsc,x_D)$. $f(\x)$ is discrete separable convex function if $f_d$ is convex\footnotemark\ for all $d$.
\end{definition}
\footnotetext{A univariate discrete function $f\colon\Z\mapsto\R_+$ is convex if $f(x+1)+f(x-1)\geq 2f(x)$ for all $x\in\Z$.}

\begin{definition}[submodularity \cite{Murota2003,Hajek1985}] \label{def:submodular}
Let $\ev_i\in{\Z^D}$ be a $D$-tuple with all zero entries except the $i$th entry being one. $f\colon\Z^D\mapsto{\R_+}$ is submodular if $f(\x+\ev_i)+f(\x+\ev_j)\geq{f(\x)+f(\x+\ev_i+\ev_j)}$ for all $\x\in{\Z^D}$ and $1 \leq i,j \leq D$.
\end{definition}

\begin{definition}[$\LN$-convexiy \cite{Murota2003}]\label{def:LConvex}
$f:\Z^{D}\mapsto{\R}_+$ is $\LN$-convex if $\psi_f(\x,\zeta)=f(\x-\zeta\One)$ is submodular in $(\x,\zeta)$, where $\One=(1,1,\dotsc,1)\in{\Z^D}$ and $\zeta\in{\Z}$.
\end{definition}

Separable convexity is the simplest case in multivariate discrete convexity, the minimization of which is easy to solve: the minimizer can be searched in $D$ directions one-by-one\cite{Murota2005}. $\LN$-convexity is defined based on the mid-point discrete convexity\cite{Murota2003}: An $\LN$-convex function $f$ satisfies
\begin{equation}
f(\x)+f(\y) \geq f \Big( \Big\lfloor{\frac{\x+\y}{2}}\Big\rfloor \Big) + f \Big( \Big\lceil{\frac{\x+\y}{2}}\Big\rceil \Big)
\end{equation}
for all $\x,\y\in\Z^D$, where $\lfloor{\x}\rfloor$ and $\lceil{\x}\rceil$ are the largest integer less than $\x$ and the smallest integer greater than $\x$, respectively.\footnote{Let $\x,\y\in\R^D$ where $\x=(x_1,\dotsc,x_D)$ and $\y=(y_1,\dotsc,y_D)$. We say that $\x\geq\y$ if $x_d\geq y_d$ for all $d\in\{1,\dotsc,D\}$.} Every discrete separable function is also $\LN$-convex\cite{Murota2005}.

\subsection{Monotonic Optimal Policy}

In this section, we show the monotonicity of optimal transmission policy $\theta^*$ in the queue state $b$.
\begin{proposition} \label{prop1}
$Q(\x,a)$ is submodular in $(b,a)$ for all $h$.
\end{proposition}
\begin{IEEEproof}
Function $Q(\x,a)$ in \eqref{eq:Q} can be rewritten as
\begin{align}
            Q(\x,a)&=Q(b,h,a) \nonumber \\
                    &=c_{tr}(h,a)+ \sum_{h'}P_{hh'} \E_f \bigg[ w\Big[ [b-a]^{+}+f-L \Big]^{+}  + \beta V(\min\{[b-a]^{+}+f,L\},h')\bigg]. \nonumber
\end{align}
Here, $Q$ is nondecreasing in $b$ and submodular in $(b,a)$ for all $V(b',h')$ that is nondecreasing and convex in $b'$ (see proof in Appendix~\ref{appAMC1}); $V(b,h)=\min_a Q(b,h,a)$ is nondecreasing and convex in $b$ for all $Q(b,h,a)$ that is nondecreasing in $b$ and submodular in $(b,a)$ (see proof in Appendix~\ref{appAMC2}). Assume that DP starts with $V^{(0)}(\x)=0$ for all $\x$. Then, by induction, Theorem~\ref{theo:Mono} holds. The optimal policy $\theta^*$ is nondecreasing in $b$ for all $h$.
\end{IEEEproof}

\begin{remark}
    Submodularity is a commonly seen property of queue departure controlled problems. One can refer to \cite{Ngo2009,Djonin2007Struct,Ding2013Mono,Huang2010} for the proofs of submodularity of $Q(\x,a)$ when different definitions of $c_q$ and $c_{tr}$ are used, e.g., $c_q=\frac{b}{\E[f]}$ as in \cite{Djonin2007Struct}.
\end{remark}

Based on Proposition~\ref{prop1}, we can prove the monotonicity of the optimal policy in queue state as follows.

\begin{theorem}  \label{theo:Mono}
The optimal policy $\theta^*$, the solution of \eqref{eq:obj}, is nondecreasing in $b$ for all $h$, i.e., $\theta^*$ is in the form of
\begin{equation} \label{eq:threshold}
    \theta^*(b,h)=\IND_{\{b\geq{\Th_h^*}\}}=
	   \begin{cases}		
		  1 & b\geq{\Th_h^*}\\
		  0 & b<\Th_h^*
	   \end{cases},
\end{equation}
where $\Th_h^*$ is the optimal queue threshold associated with channel state $h$.
\end{theorem}
\begin{IEEEproof}
We use the following property of submodular functions\cite{Topkis1978}: If $Q$ is submodular in $(b,a)$ for all $h$, the minimizer $a^*(\x)=\arg\min_a Q(\x,a)$ is nondecreasing in $b$ for all $h$. According to Proposition~\ref{prop1}, $Q(\x,a) = c(\x,a)+\beta\sum_{\x'}P_{\x\x'}^{a}V^{(N)}(\x')$ is submodular in $(b,a)$ for all $h$. Therefore, $\theta^*$ is nondecreasing in $b$. Theorem holds.
\end{IEEEproof}

\begin{figure}[tb]
	\centering
        \subfigure[the optimal policy $\theta^*$ as a function of queue state $b$ and channel state $h$]{\scalebox{0.7}{\input{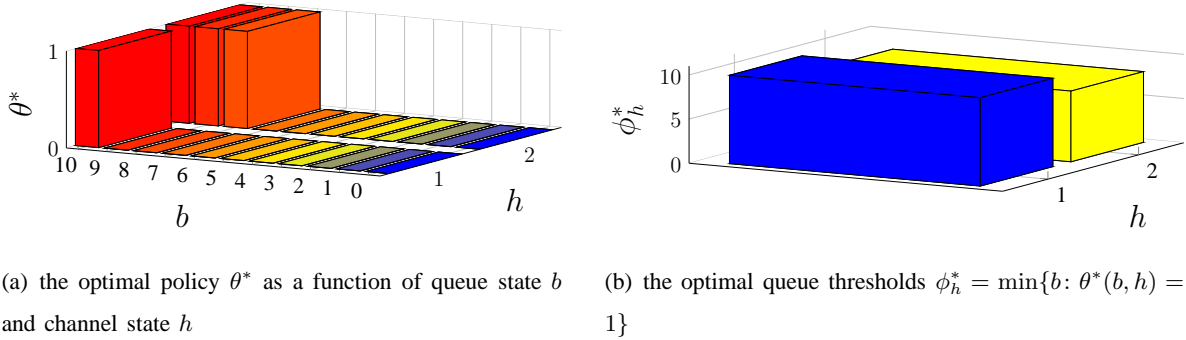}}} \quad
        \subfigure[the optimal queue thresholds $\Th_h^*=\min\{b \colon \theta^*(b,h)=1\}$]{\scalebox{0.7}{
%
%
\begin{tikzpicture}

\begin{axis}[%
width=3.7in,
height=1.3in,
unbounded coords=jump,
view={-150}{30},
scale only axis,
xmin=0.5,
xmax=1.5,
xtick={\empty},
xmajorgrids,
y dir=reverse,
ymin=0.5,
ymax=2.5,
ytick={1,2},
yticklabels={1,2},
ylabel={\Large $h$},
ymajorgrids,
zmin=0,
zmax=11,
zlabel={\Large $\Th_h^*$},
zmajorgrids,
axis x line*=bottom,
axis y line*=left,
axis z line*=left
]

\addplot3[%
surf,
shader=flat,
colormap={summap}{
color=(yellow); color=(blue);
color=(black); color=(white)
color=(orange) color=(violet)
color=(red)
},
draw=black,
point meta=explicit,
mesh/rows=4]
table[row sep=crcr,header=false,meta index=3] {
NaN NaN NaN 1\\
0.6 1.6 0 1\\
0.6 2.4 0 1\\
NaN NaN NaN 1\\
NaN NaN NaN 1\\
NaN NaN NaN 1\\
0.6 1.6 0 1\\
0.6 1.6 8 1\\
0.6 2.4 8 1\\
0.6 2.4 0 1\\
0.6 1.6 0 1\\
NaN NaN NaN 1\\
1.4 1.6 0 1\\
1.4 1.6 8 1\\
1.4 2.4 8 1\\
1.4 2.4 0 1\\
1.4 1.6 0 1\\
NaN NaN NaN 1\\
NaN NaN NaN 1\\
1.4 1.6 0 1\\
1.4 2.4 0 1\\
NaN NaN NaN 1\\
NaN NaN NaN 1\\
NaN NaN NaN 1\\
};

\addplot3[%
surf,
shader=flat,
colormap={summap}{
color=(blue); color=(blue);
color=(black); color=(white)
color=(orange) color=(violet)
color=(red)
},
draw=black,
point meta=explicit,
mesh/rows=4]
table[row sep=crcr,header=false,meta index=3] {
NaN NaN NaN 1\\
0.6 0.6 0 1\\
0.6 1.4 0 1\\
NaN NaN NaN 1\\
NaN NaN NaN 1\\
NaN NaN NaN 1\\
0.6 0.6 0 1\\
0.6 0.6 10 1\\
0.6 1.4 10 1\\
0.6 1.4 0 1\\
0.6 0.6 0 1\\
NaN NaN NaN 1\\
1.4 0.6 0 1\\
1.4 0.6 10 1\\
1.4 1.4 10 1\\
1.4 1.4 0 1\\
1.4 0.6 0 1\\
NaN NaN NaN 1\\
NaN NaN NaN 1\\
1.4 0.6 0 1\\
1.4 1.4 0 1\\
NaN NaN NaN 1\\
NaN NaN NaN 1\\
NaN NaN NaN 1\\
};

\end{axis}
\end{tikzpicture}%
}}
	\caption{The optimal policy and queue threshold vector in a single-user system (Fig.~\ref{fig:QG}), where $w=4$, $f^{(t)}\sim{\text{Bernoulli}(0.5)}$ for all $t$, $\bar{P_b}=0.01$ and $L=10$. The channel is modeled by a $2$-state FSMC.}
	\label{fig:AMC_P}
\end{figure}

\subsection{Discrete Convex Minimization Problem}
\label{sec:ProblemForm}

Let $\Th_h^*=\min\{b \colon \theta^*(\x)=1\}$. It follows that the optimal monotonic policy $\theta^*$ is fully characterized by the optimal queue thresholds $\Th_h^*$ for all $h$ if Theorem~\ref{theo:Mono} holds. There is an example of optimal queue threshold $\Th_h^*$ in Fig.~\ref{fig:AMC_P}. Let $\theta$ ba a deterministic policy that is nondecreasing in $b$. Define a threshold vector $\Thv=(\Th_1,\Th_2,\dotsc,\Th_{|\HSet|})$, where $\Th_h=\min\{b\colon\theta(b,h)=1\}\in\B$. We show in the following theorem that \eqref{eq:obj} can be converted to a queue threshold vector optimization problem with a discrete convex objective function.

\begin{theorem} \label{theo:Lconvex}
Let $\ThSet=\B^{|\HSet|}$. If Theorem~\ref{theo:Mono} holds, then \eqref{eq:obj} is equivalent to
\begin{equation}\label{eq:LConvexOpt}
        \min_{ \Thv \in \ThSet} J(\Thv),
\end{equation}
where the objective function
\begin{equation} \label{eq:J}
    J(\Thv)=\sum_{\x} \E \bigg[ \sum_{t=0}^{\infty} \beta^t c(\x^{(t)},\IND_{\{b^{(t)}\geq{\Th_{h^{(t)}}}\}}) \Big| \x^{(0)}=\x \bigg]
\end{equation}
is both discrete separable convex and $\LN$-convex in $\Thv$.
\end{theorem}
\begin{IEEEproof}
Let $\theta$ be the policy determined by $\Thv$ through $\theta(b,h)=\IND_{\{b\geq{\Th_h}\}}$. According to \eqref{eq:Vtheta}, we have $J(\Thv)=\sum_{\x}V_{\theta}(\x)$. Therefore, \eqref{eq:obj} is equivalent to $\min_{\Thv}J(\Thv)$. Define $V_b(h,\Th_h)=\sum_{b}Q(b,h,\IND_{\{b\geq{\Th_h}\}})$. Due to the submodularity of $Q$ in $(b,a)$, we have
    \begin{align}\label{eq:IntegerConvexOfVb}
        &\quad V_b(h,\Th_h+1)+V_b(h,\Th_h-1)-2V_b(h,\Th_h)                   \nonumber \\
        &=Q(\Th_h,h,0)-Q(\Th_h-1,h,0)+Q(\Th_h-1,h,1) -Q(\Th_h,h,1) \geq{0}.
    \end{align}
    So, $V_b$ is convex in $\Th_h$ for all $h$. Since $J$ can be expressed in the form of
    \begin{align}
        J(\Thv) &=\sum_{h}\sum_{b}Q(b,h,\IND_{\{b\geq{\Th_h}\}})    \nonumber \\
                &=\sum_{h}V_b(h,\Th_h).
    \end{align}
By Definition~\ref{def:DiscSep}, $J$ is discrete separable convex in $\Thv$. Since every discrete separable convex function is $\LN$-convex, $J$ is also $\LN$-convex in $\Thv$.
\end{IEEEproof}

Problem~\eqref{eq:LConvexOpt} is different from the conventional convex optimization problems. Firstly, \eqref{eq:LConvexOpt} is an integer programming, or discrete optimization, problem where most of the techniques designed for continuous optimization may not be directly applicable. Secondly, the objective function $J$ in \eqref{eq:LConvexOpt} is an expectation, i.e., \eqref{eq:LConvexOpt} is a stochastic optimization problem rather than a deterministic one. Therefore, we consider DSA algorithms, the SA algorithms that is exclusively proposed for discrete stochastic minimization problems, for solving \eqref{eq:LConvexOpt} in the next section.

\section{Discrete Stochastic Approximation}
\label{sec:SA}

This section focuses on two DSA algorithms, DSPSA\cite{Wang2011} and $\LN$-convex SA\cite{LimDSA2012}, for solving problem \eqref{eq:LConvexOpt}. They are specifically designed for discrete convex minimization problems where almost sure convergence performance is achievable. But, it should be pointed out that the solution to problem \eqref{eq:LConvexOpt} is not restricted to DSA methods. With the objective function being discrete convex, there may exist many methods that converge with probability $1$ \cite{Andradottir:1999}, e.g., random search\cite{Rastrigin1964}, simulated annealing\cite{Kirkpatrick1984}. This paper considers two such methods where the conditions for almost sure convergence for problem \eqref{eq:LConvexOpt} are straightforwardly satisfied.

	\begin{algorithm} [t]
	\label{algo:DSA}
	\small
	\SetAlgoLined
	\SetKwInOut{Input}{input}\SetKwInOut{Output}{output}
	\SetKwFor{For}{for}{do}{endfor}
	\Input{initial guess $\tilde{\Thv}^{(0)}$ (a $D$-tuple), total number of iterations $N$, step size parameters $A$, $B$ and $\alpha$}
	\Output{$[\tilde{\Thv}^{(N)}]$, the closest integer point to $\tilde{\Thv}^{(N)}$ by Euclidean distance. }
	\BlankLine
	\Begin{
        \For {n=1 \emph{\KwTo} N} {
            $a^{(n)}=\frac{A}{(B+n)^{\alpha}}$\;
            obtain $\gv$ at $\tilde{\Thv}^{(n-1)}$ by using $\hat{J}$\;
            $\tilde{\Thv}^{(n)}=\tilde{\Thv}^{(n-1)}-a^{(n)}\gv(\tilde{\Thv}^{(n-1)})$\;
        }
	}
	\caption{DSA \cite{Wang2011,LimDSA2012}}
	\end{algorithm}

\begin{figure*}[t]
	\centering
        \subfigure[{an $\LN$-convex function $f \colon \{0,\dotsc,3\}^2\mapsto\R_+$}]{\includegraphics[height=3.5cm,width=8.5cm]{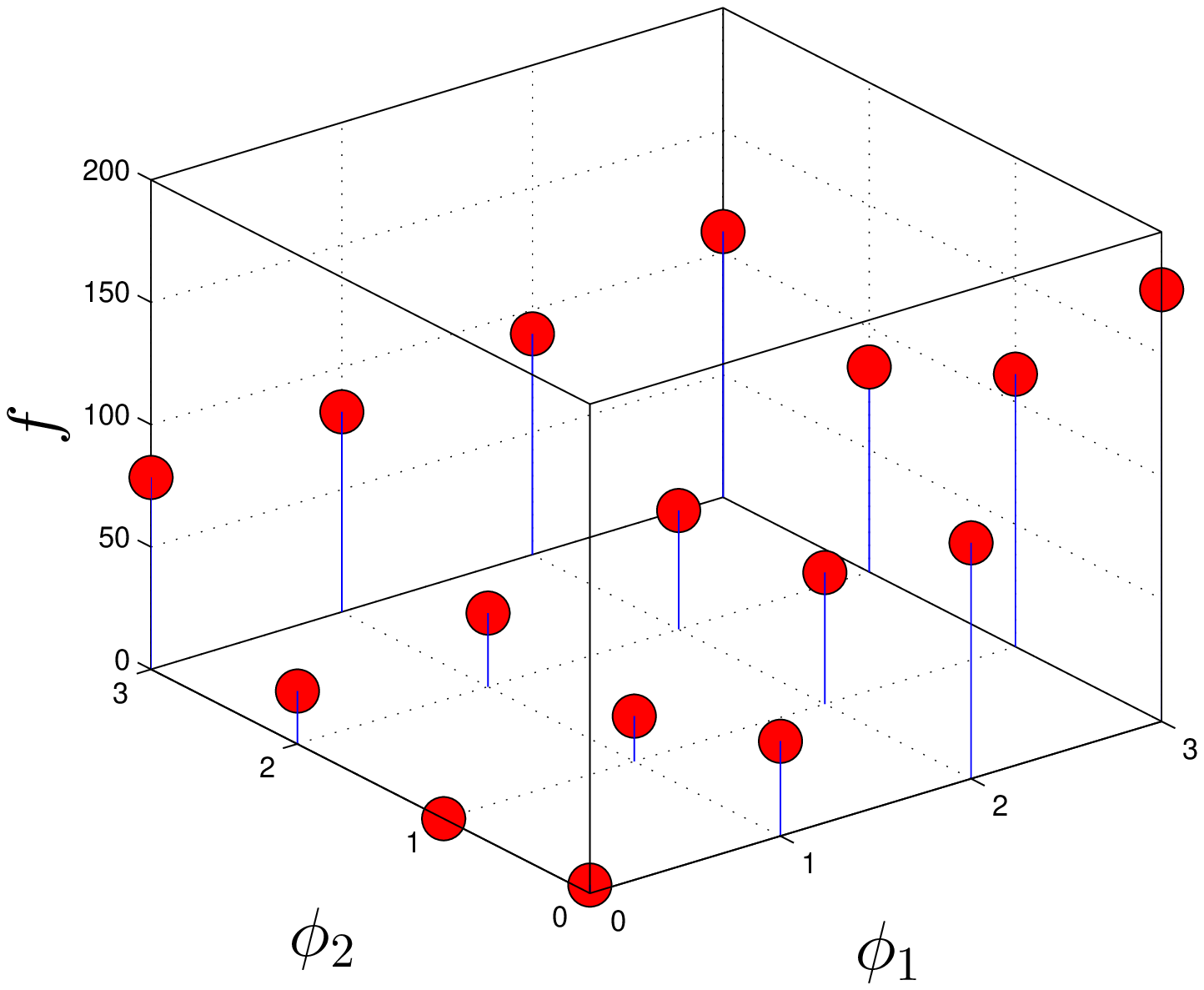}} \qquad
        \subfigure[{the PLI function $\tilde{f} \colon [0,3]^2\mapsto\R_+$}]{\includegraphics[height=3.5cm,width=8.5cm]{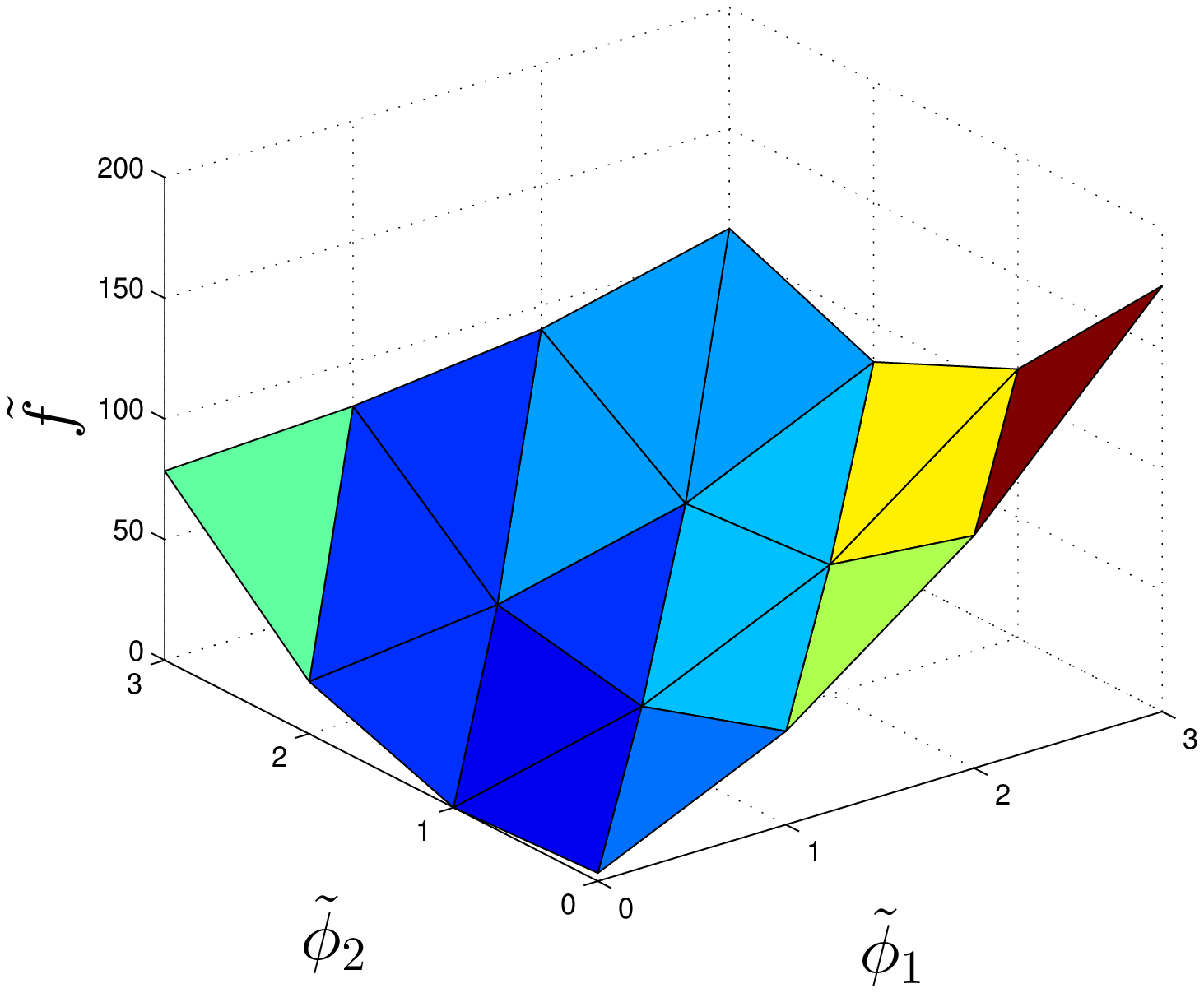}}\qquad
	\caption{An example of PLI function. {Let $\Thv=(\Th_1,\Th_2)\in\{0,\dotsc,3\}^2$ and $\tilde{\Thv}=(\tilde{\Th}_1,\tilde{\Th}_2)\in[0,3]^2$.} According to \cite{Murota2003,LimDSA2012}, $\min_{\Thv}f(\Thv)=\min_{\tilde{\Thv}}\tilde{f}(\tilde{\Thv})$ and $\arg\min_{\Thv}f(\Thv)=\arg\min_{\tilde{\Thv}}\tilde{f}(\tilde{\Thv})$. }
	\label{fig:Lconvex}
\end{figure*}

Both DSPSA and $\LN$-convex SA are based on a line search method. They assume that a noisy measurement of $J$,
\begin{equation}
    \hat{J}(\Thv)=J(\Thv)+z,
\end{equation}
is obtainable. Here, $z$ is the random measurement noise. They follow the procedure of DSA algorithm as shown in Algorithm 1. Each of them generates a sequence of estimations $\{\tilde\Thv{}^{(n)}\}$ by a line search iteration
\begin{equation}
    \tilde{\Thv}\colonequals \tilde{\Thv}-a\gv(\tilde{\Thv}).
\end{equation}
For problem \eqref{eq:LConvexOpt}, $D$ is the dimension of $\tilde{\Thv}^{(n)}$, and $\tilde{\Thv}^{(n)}\in\tilde{\ThSet}=[0,L]^{D}$. The two algorithms differ in how they obtain the gradient $\gv(\tilde{\Thv}^{(n)})$.

\subsection{Discrete Simultaneous Perturbation Stochastic Approximation\cite{Wang2011}}
Let $\mathbf{\Delta}=(\Delta_1,\dotsc,\Delta_D)$ with each tuple $\Delta_d\in\{-1,1\}$ being independent Bernoulli random variables with probability $0.5$. The $d$th entry of $\gv(\tilde{\Thv}^{(n)})$ is obtained by
\begin{equation} \label{eq:DSPSA}
    g_d(\tilde{\Thv}^{(n)})= \bigg[ \hat{J} \Big( \lfloor\tilde{\Thv}^{(n)}\rfloor+\frac{\One+\mathbf{\Delta}}{2} \Big)  - \hat{J} \Big( \lfloor\tilde{\Thv}^{(n)}\rfloor+\frac{\One-\mathbf{\Delta}}{2} \Big) \bigg] \Delta_d^{-1}.
\end{equation}
As explained in \cite{Wang2011}, $\gv$ is obtained as the gradient based on the discrete mid-point convexity. For separable discrete convex minimization problem, the sequence $\{\tilde{\Thv}^{(n)}\}$ converges almost surely if the standard conditions\footnote{The standard conditions are $a^{(n)}>0,\lim_{n\rightarrow\infty}a^{(n)}=0,\sum_n a^{(n)}=\infty,\sum_n (a^{(n)})^2<\infty$ and $z$ has zero mean and uniformly bounded variance.} are satisfied\cite{Wang2011}.

\subsection{$\LN$-convex Stochastic Approximation\cite{LimDSA2012}}

$\LN$-convex SA is in fact applied to the piecewise linear interpolation (PLI) of the discrete objective function. The PLI of an $\LN$-convex function is defined as follows.

Let $\tilde{\Thv}\in\tilde{\ThSet}^{\phantom{a}}$. Denote $\mathbf{p}=\lfloor{\tilde{\Thv}}\rfloor$, $\mathbf{q}=\tilde{\Thv}-\mathbf{p}$ and
\begin{equation}
    U_d=\begin{cases} \emptyset, & d=0 \\ \{\sigma(1),\dotsc,\sigma(d)\}, & d\neq{0} \end{cases},
\end{equation}
where $\sigma$ is the permutation of $(1,\dotsc,D)$ such that $\sigma(d)$ is the index of the $d$th largest of $q_1,\dotsc,q_D$, the components in $\mathbf{q}$. Let $\Charact\in\{0,1\}^D$ be a characteristic vector whose $d$th entry is $1$ when $d$ belongs to $U_d$ and $0$ otherwise. If $J$ is a discrete function in $\Thv$, its PLI function $\tilde{J}$ is defined by
\begin{align}
    \tilde{J}(\Thv)=&(1-q_{\sigma(1)})J(\mathbf{p})+(q_{\sigma(1)}-q_{\sigma(2)})J(\mathbf{p}+\chi_{\scalebox{0.6}{$U_1$}}^{\phantom{l}})\dotsc  \nonumber \\
                    &+q_{\sigma(D)}J(\mathbf{p}+\chi_{\scalebox{0.6}{$U_D$}}^{\phantom{l}}).
\end{align}
If $J$ is an $\LN$-convex function in $\Thv$, $\tilde{J}$ is a continuous convex function in $\tilde{\Thv}$, and the minimizers and minima of $\tilde{J}$ agree with those of $J$ \cite{Murota2005} (See Fig.~\ref{fig:Lconvex} for an example). Therefore, the minimizers of $J$ can be approximated by a line search algorithm applied to $\tilde{J}$. In $\LN$-convex SA, $\gv(\tilde{\Thv}^{(n)})$ is obtained as a subgradient\footnotemark of $\tilde{J}$. This subgradient is calculated by using the noisy measurements $\hat{J}$ as follows.
\footnotetext{$\rho(\x)$ is called the subgradient of $f$ at $\x$ if $f(\y)-f(\x)\geq \rho(\x)(\y-\x)$\cite{Rockafellar1997}. For a nonsmooth function, there may be more than one subgradient at $\x$. The work in \cite{LimDSA2012} shows how to calculate one such subgradient.}

Define $Y(d)$ such that
    \begin{align} \label{eq:subgrad1}
        Y(0)&=\hat{J}(\mathbf{p}^{(n)}),                                                           \nonumber \\
        Y(d)&=\hat{J}(\mathbf{p}^{(n)}+\Charact),
    \end{align}
where $\mathbf{p}^{(n)}=\lfloor{\tilde{\Thv}^{(n)}}\rfloor$ and $\Charact$ is obtained by using $\mathbf{q}^{(n)}=\tilde{\Thv}^{(n)}-\mathbf{p}^{(n)}$. The $d$th entry of subgradient $\gv$ at $\tilde{\Thv}^{(n)}$ is
    \begin{align} \label{eq:subgrad2}
        g_d(\tilde{\Thv}^{(n)})=Y(\sigma(d))-Y(\sigma(d)-1).
    \end{align}

Unlike DSPSA, $\LN$-convex SA does not using random perturbations to estimate $\gv$. Instead, it uses $D+1$ measurements of $\hat{J}$ to get more accurate estimate of the descent direction. If the standard conditions are satisfied, the sequence $\{\tilde{\Thv}^{(n)}\}$ converges almost surely for $\LN$-convex minimization problems\cite{LimDSA2012}. It is also shown in \cite{LimRateOfC2011} that $\{\tilde{\Thv}^{(n)}\}$ converges with a rate of $1/n$ on average.

\subsection{Implementation of DSA Algorithms}
We list below the implement details when we apply two DSA methods to produce the results in Section~\ref{sec:NumRes}.
\subsubsection{Step Size}
The step size parameters, $A$, $B$ and $\alpha$, in Algorithm 1 are crucial for the convergence performance of DSA algorithms. As aforementioned, they must be chosen to satisfy the standard conditions. We adopt the method of choosing $A$, $B$ and $\alpha$ suggested in \cite{SpallPar1998} for practical problems where the computation budget $N$, the total number of iterations, is fixed: $B=0.095N$, $\alpha=0.602$ and $A$ is chosen so that $A/(B+1)^{\alpha}\|\gv(\tilde{\Thv}^{(0)})\|$ achieves the desired change of $\tilde{\Thv}^{(1)}$. In all the experiments in Section~\ref{sec:NumRes}, we assign $\tilde{\Thv}^{(0)}=\Zero$ and $N=500$. Therefore, $B$ is fixed to $47.5$. We assume the desired value of $A/(B+1)^{\alpha}\|\gv(\tilde{\Thv}^{(0)})\|$ is $0.1$. Before each time we implement DSPSA or $\LN$-convex SA, we run $100$ repetitions to obtain a reliable estimation of $\|\gv(\tilde{\Thv}^{(0)})\|$ (the value averaged over repetitions) and then select $A$ such that $A/(B+1)^{\alpha}\|\gv(\tilde{\Thv}^{(0)})\|=0.1$. Since $\|\gv(\tilde{\Thv}^{(0)})\|$ varies with each system and DSA method, we show the value of $A$ for each experiment in Section~\ref{sec:NumRes}.

\subsubsection{Obtaining $\hat{J}$}
The method of obtaining $\hat{J}$ at $\Thv$ is to simulate the state sequence $\{\x^{(t)}\}$. Here, $\x^{(t)}$ varies according to the Markov chain that is governed by the transition probability $\Pr(\x^{(t+1)}|\x^{(t)})=P_{\x^{(t)}\x^{(t+1)}}^{\theta(\x^{(t)})}$, where $\theta(\x)=\IND_{\{b\geq{\Th_{h}}\}}$. We obtain $\hat{J}$ as
\begin{align} \label{eq:SimJ}
    \hat{J}(\Thv)=\frac{1}{N_{r}}\sum_{\x^{(0)}\in\X}\sum_{i=1}^{N_{r}}\sum_{t=0}^{T}\beta^{t}c(\x^{(t)},\IND_{\{b^{(t)}\geq{\Th_h}\}}),
\end{align}
i.e., $\hat{J}$ is the value averaged over $N_r$ simulations.\footnote{Most SPSA algorithms just require a single simulation to obtain $\hat{J}(\Thv)$. We use repetition because the average value of multiple simulations was suggested in \cite{Spall1992mult,LimDSA2012} to improve the convergence performance.} We fix $N_r$ to $100$. The simulation length $T$ depends on $\beta$, i.e., the simulation stops until the increments over several successive decision epochs is blow a small threshold ($10^{-4}$). In this paper, $\beta$ is fixed to $0.95$.

\section{Numerical Results}
\label{sec:NumRes}
In this section, we run experiments in three cross-layer on-off transmission control systems, one single-user and two multi-user systems. In each system, we implement two DSA algorithms, DSPSA and $\LN$-convex SA. Their convergence performances are compared to a CSPSA algorithm.

The CSPSA algorithm is an SA algorithm that is originally proposed for continuous stochastic minimization problems. It follows the same procedure as in Algorithm 1. It uses the same perturbation vector $\mathbf{\Delta}$ as in DSPSA to obtain the gradient $\gv$. But, the $d$th entry of $\gv(\tilde{\Thv}^{(n)})$ is given by
\begin{equation}
    g_d(\tilde{\Thv}^{(n)})=\frac{\hat{J}(\Gamma(\tilde{\Thv}^{(n)}+\Delta_d))-\hat{J}(\Gamma(\tilde{\Thv}^{(n)}-\Delta_d))}{2c^{(n)}\Delta_d}, \nonumber
\end{equation}
where $c^{(n)}=\frac{C}{n^{\rho}}$ and $\Gamma$ is the projection function proposed in \cite{Bhatnagar2011} and is given by
\begin{equation} \label{eq:project}
    \Gamma(\tilde{\Thv})=\begin{cases}
        \lfloor{\tilde{\Thv}}\rfloor & \text{w/\ prob.}\ \frac{\lceil{\tilde{\Thv}}\rceil-\tilde{\Thv}}{\lceil{\tilde{\Thv}}\rceil-\lfloor{\tilde{\Thv}}\rfloor} \\
        \lceil{\tilde{\Thv}}\rceil & \text{w/\ prob.}\ \frac{\tilde{\Thv}-\lfloor{\tilde{\Thv}}\rfloor}{\lceil{\tilde{\Thv}}\rceil-\lfloor{\tilde{\Thv}}\rfloor}
        \end{cases}.
\end{equation}
The method to implement $\Gamma(\tilde{\Thv})$ is: The scheduler adopts $\lfloor{\tilde{\Thv}}\rfloor$ sometimes and $\lceil{\tilde{\Thv}}\rceil$ the other times so that in the long run it chooses policy $\lfloor{\tilde{\Thv}}\rfloor$ with probability $\frac{\lceil{\tilde{\Thv}}\rceil-\tilde{\Thv}}{\lceil{\tilde{\Thv}}\rceil-\lfloor{\tilde{\Thv}}\rfloor}$ and $\lceil{\tilde{\Thv}}\rceil$ with probability $\frac{\tilde{\Thv}-\lfloor{\tilde{\Thv}}\rfloor}{\lceil{\tilde{\Thv}}\rceil-\lfloor{\tilde{\Thv}}\rfloor}$. The step size parameters for CSPSA are $A$, $B$, $\alpha$, $C$ and $\rho$. They are also chosen by following the suggestion in \cite{SpallPar1998}. We set $B=0.095N$, $\alpha=0.602$, $C=1$ and $\rho=0.101$. $A$ is chosen so that $A/(B+1)^{\alpha}\|\gv(\tilde{\Thv}^{(0)})\|=0.1$. The value of $A$ is given in each experiment.

We also run DP to obtain the optimal policy $\theta^*$. $\Thv^*$, the optimal threshold vector, and $J(\Thv^*)$, the minimum of \eqref{eq:LConvexOpt}, are calculated by using \eqref{eq:threshold} and \eqref{eq:J}, respectively. We show the convergence performance in terms of the following two metrics:
\begin{itemize}
    \item $J([\tilde{\Thv}^{(n)}])$: the value of the objective function at $[\tilde{\Thv}^{(n)}]$, the closest integer point to $\tilde{\Thv}^{(n)}$;
    \item $\frac{\|\tilde{\Thv}^{(n)}-\Thv^*\|}{\|\tilde{\Thv}^{(0)}-\Thv^*\|}$: the normalized error of the estimation $\tilde{\Thv}^{(n)}$.
\end{itemize}

\subsection{Single-user System}
\label{sec:singleuser}

\begin{figure}[tb]
	\centering
        \subfigure[{the value of the objective function $J([\tilde{\Thv}^{(n)}])$ vs. iteration}]{\scalebox{0.85}{\input{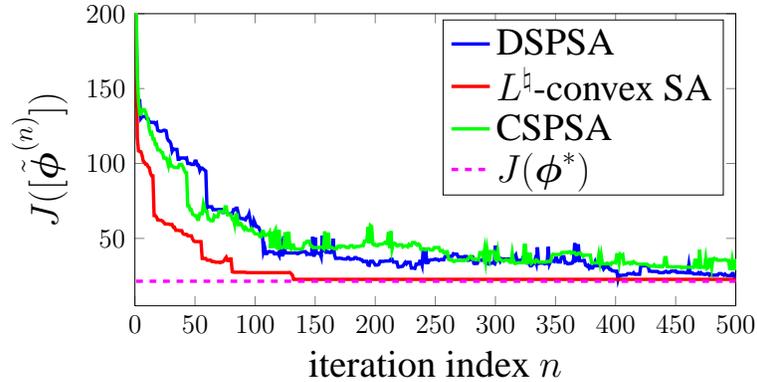}}} \quad
        \subfigure[the normalized error $\frac{\|\tilde{\Thv}^{(n)}-\Thv^*\|}{\|\tilde{\Thv}^{(0)}-\Thv^*\|}$ vs. iteration]{\scalebox{0.85}{\input{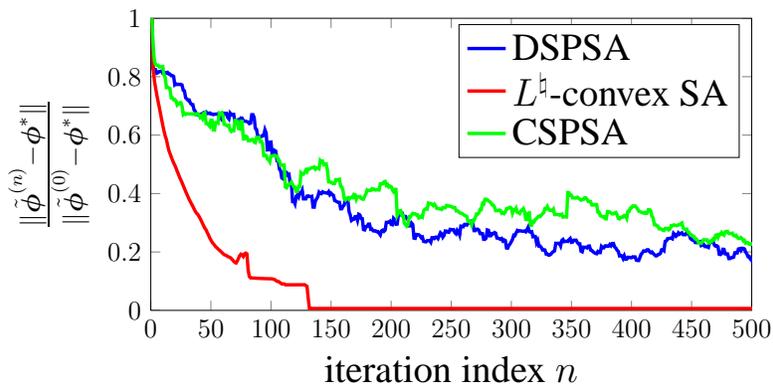}}}
	\caption{Convergence performances of DSPSA, $\LN$-convex SA and CSPSA in a single-user system, where channel is modeled by a $8$-state FSMC. $w=4$, $f^{(t)}\sim{\text{Bernoulli}(0.5)}$ for all $t$, $\bar{P_b}=0.01$ and $L=10$. The dimension of $\Thv$ is $8$.}
	\label{fig:AMC2}
\end{figure}

Consider the on-off transmission control system in Fig.~\ref{fig:QG}. We set $w=4$, $L=10$, $p_f^{\phantom{a}}=0.5$ and $\bar{P_b}=0.01$. Let the channel experience slow and flat Rayleigh fading with average SNR being $0\si{\decibel}$ and maximum doppler shift being $10\text{Hz}$. We adopt an $8$-state FSMC model. In this experiment, the step size parameter $A$ is $0.75$ for DSPSA, $0.9$ for $\LN$-convex SA and $0.7$ for CSPSA. The results are shown in Fig.~\ref{fig:AMC2}. It can be seen that the convergence performance of DSPSA is comparable to that of CSPSA. $\LN$-convex SA has $\{J([\tilde{\Thv}^{(n)}])\}$ converges faster than DSPSA and CSPSA.

\subsection{Multi-user Systems}
\label{sec:multiuser}
In a multi-user system, we denote $i$ the user index. The packets sent from user $i$ is buffered by a queue (We call it queue $i$), and the departure packets of queue $i$ are transmitted through channel $i$. We use subscript $i$ to denote the variable associated to user $i$. Let $L_i$ be the length of queue $i$. We assume all queues has the same length, i.e., $L_i=L$ for all $i$. The action $a_i\in\{0,1\}$ determines the number of packets departs from queue $i$. We assume that the Assumptions~\ref{ass1} to \ref{ass3} hold for each user. In this section, we run experiments in a five-user orthogonal frequency-division multiple-access (OFDMA) system and a four-user network-coded two-way relay channel (NC-TWRC) system. In both systems, we set $L=5$, $w=4$ and $\bar{P_b}=0.01$.

\begin{figure}[tpb]
	\centering
		\scalebox{0.8}{\begin{tikzpicture}

\draw [ ->] (0.5,0) -- (1.5,0);
\node at (1,0.25) {$f_1^{(t)}$};

\node at (4.5,0.8) {\small scheduler};
\draw (3.6,0.6) rectangle (5.4,1);
\draw [dashed,->] (4.5,0.6) -- (4.5,0.18);
\draw [dashed,->] (4.4,0.6) -- (4.4,-0.8) -- (4.5,-0.8) -- (4.5,-1);
\draw [dashed,->] (4.2,0.6) -- (4.2,-2.6) -- (4.5,-2.6) -- (4.5,-2.8);

\draw [fill=red!20] (2.8,0.35) rectangle (3.4,-0.35);
\draw (1.9,0.35) rectangle (3.4,-0.35);
\draw (1.9,0.35) -- (1.5,0.35);
\draw (1.9,-0.35) -- (1.5,-0.35);
\draw (2.2,0.35) -- (2.2,-0.35);
\draw (2.5,0.35) -- (2.5,-0.35);
\draw (2.8,0.35) -- (2.8,-0.35);
\draw (3.1,0.35) -- (3.1,-0.35);
\node at (0.6,-0.4) {\small user $1$};

\draw (3.4,0) -- (4.3,0);
\draw (4.3,0) -- (4.8,0.3);
\draw [->] (4.8,0) -- (5.8,0) -- (5.8,-1)--(6.5,-1);

\node at (3.8,-1.8) {$\vdots$};

\draw [ ->] (0.5,-1.2) -- (1.5,-1.2);
\node at (1,-0.95) {$f_2^{(t)}$};

\draw [fill=blue!20] (2.8,-0.85) rectangle (3.4,-1.55);
\draw (1.9,-0.85) rectangle (3.4,-1.55);
\draw (1.9,-0.85) -- (1.5,-0.85);
\draw (1.9,-1.55) -- (1.5,-1.55);
\draw (2.2,-0.85) -- (2.2,-1.55);
\draw (2.5,-0.85) -- (2.5,-1.55);
\draw (2.8,-0.85) -- (2.8,-1.55);
\draw (3.1,-0.85) -- (3.1,-1.55);
\node at (0.6,-1.6) {\small user $2$};

\draw (3.4,-1.2) -- (4.3,-1.2);
\draw (4.3,-1.2) -- (4.8,-0.9);
\draw [->] (4.8,-1.2) -- (5.8,-1.2) -- (5.8,-1.2)--(6.5,-1.2);

\draw [ ->] (0.5,-3) -- (1.5,-3);
\node at (1,-2.75) {$f_5^{(t)}$};

\draw [fill=green!20] (2.8,-2.65) rectangle (3.4,-3.35);
\draw (1.9,-2.65) rectangle (3.4,-3.35);
\draw (1.9,-2.65) -- (1.5,-2.65);
\draw (1.9,-3.35) -- (1.5,-3.35);
\draw (2.2,-2.65) -- (2.2,-3.35);
\draw (2.5,-2.65) -- (2.5,-3.35);
\draw (2.8,-2.65) -- (2.8,-3.35);
\draw (3.1,-2.65) -- (3.1,-3.35);
\node at (0.6,-3.55) {\small user $5$};

\draw (3.4,-3) -- (4.3,-3);
\draw (4.3,-3) -- (4.8,-2.7);
\draw [->] (4.8,-3) -- (5.8,-3) -- (5.8,-1.8) -- (6.5,-1.8);

\draw (6.5,-0.8) rectangle (8.2,-2);
\node at (7.35,-1.2) {\small OFDM};
\node at (7.35,-1.6) {\small transmitter};

\draw [->] (8.2,-1.5) -- (8.8,-1.5);

\node at (9.4,-1.2) {\small wireless};
\node at (9.4,-1.6) {\small channel};

\end{tikzpicture}}
	\caption{There are five users in the OFDMA system. Each user is assigned a queue and a subcarrier. The packets sent from user $i$ is buffered by the queue. The departing packets from all queues are modulated by BPSK symbols and transmitted by the orthogonal frequency-division multiplexing (OFDM) transmitter}
	\label{fig:CongestG}
\end{figure}
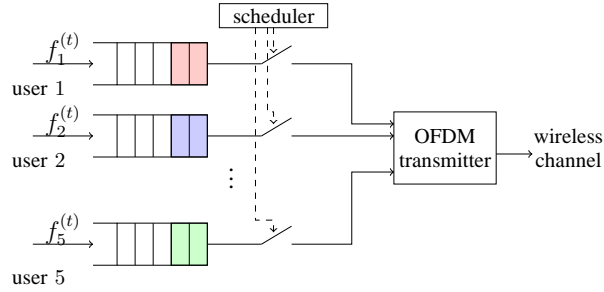

\subsubsection{Five-user OFDM System}
\label{sec:multiuser1}

Consider the OFDMA system as shown in Fig.~\ref{fig:CongestG}.\footnote{This system is a special case of the OFDMA system proposed \cite{Niyato2006}. The difference is that the problem in \cite{Niyato2006} is a variable rate adaptation one, while, in this paper, the scheduler is restricted to choose only transmit or not. } There are five users in this system. The scheduler assigns each user a subcarrier. The departing packets from queues are modulated by BPSK symbols and transmitted by the orthogonal frequency-division multiplexing (OFDM) transmitter. The departures of all queues are controlled by one scheduler. We assume $p_{f_1}=0.2$, $p_{f_2}=0.4$ and $p_{f_i}=0.5$ for all $i\in\{3,4,5\}$. In this system, channel $i$ denotes the subcarrier $i$, i.e., $\snr_i^{(t)}$ denotes the instantaneous SNR of subcarrier $i$. We assume that each channel has the average SNR $0\si{\decibel}$ and is modeled by a 4-state FSMC.

\begin{figure}[tb]
	\centering
        \subfigure[{the value of the objective function $J([\tilde{\Thv}^{(n)}])$ vs. iteration}]{\scalebox{0.85}{\input{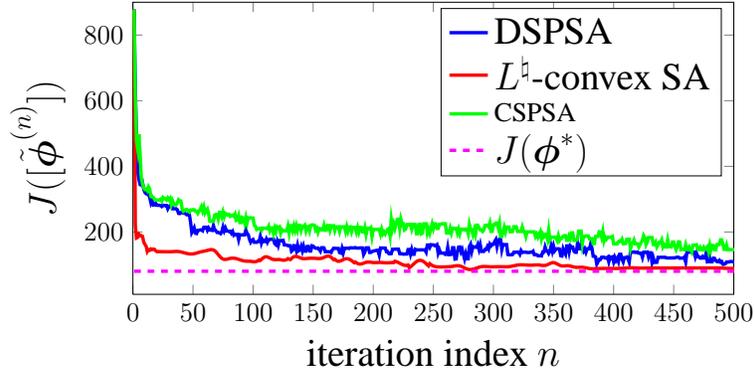}}} \quad
        \subfigure[the normalized error $\frac{\|\tilde{\Thv}^{(n)}-\Thv^*\|}{\|\tilde{\Thv}^{(0)}-\Thv^*\|}$ vs. iteration]{\scalebox{0.85}{\input{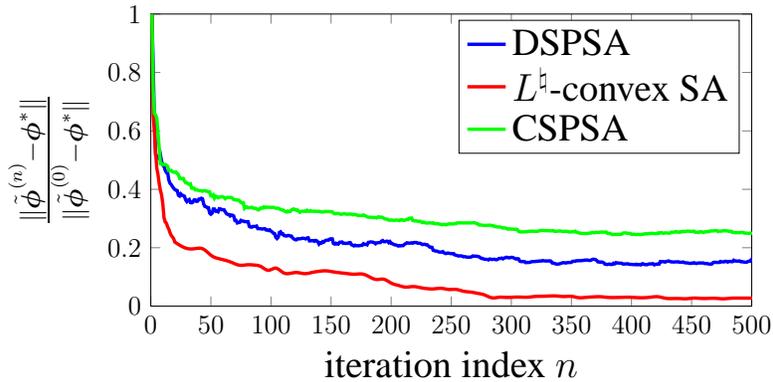}}}
	\caption{Convergence performances of DSPSA, $\LN$-convex SA and CSPSA in a five-user OFDMA system (in Fig.~\ref{fig:CongestG}). Each channel/subcarrier is modeled by a $4$-state FSMC. In this system, $L_i=5$ for all $i\in\{1,\dotsc,5\}$, $p_{f_1}=0.2$, $p_{f_2}=0.4$ and $p_{f_3}=p_{f_4}=p_{f_5}=0.5$. The dimension of the threshold vector $\Thv$ is $40$.}
	\label{fig:CongestGBR}
\end{figure}

In this system, we can formulate an MDP model as in Section~\ref{sec:MDP} for each user. For example, the MDP model for user $i$ has the state $\x=(b_i,h_i)$, action $a=a_i\in\{0,1\}$ and the state transition probability and the immediate cost are the same as described in Section~\ref{sec:MDP}. There is an optimal policy $\theta_i^*$ to each MDP model. $\theta_i^*(b_i,h_i)$ determines an optimal action to queue $i$ for state $(b_i,h_i)$. It can be seen that Theorem~\ref{theo:Mono} holds for all MDPs. Therefore, $\theta_i^*(b_i,h_i)$ is nondecreasing in $b_i$ for all $i$. Let the optimal policy in the entire system be $\theta^*=(\theta_1^*,\dotsc,\theta_5^*)$. We construct the threshold vector as follows. Let $\Thv=(\Thv_1,\dotsc,\Thv_5)$ where $\Thv_i$ is the queue threshold vector of user $i$ and is constructed by stacking $\Th_{h_i}=\min\{b_i \colon \theta_i(b_i,h_i)=1 \}$ for all $h_i$. $\hat{J}(\Thv)$ is the simulation value of the objective that is summed over all users.\footnote{The idea is to simulate $\hat{J}_i(\Thv_i)$ for all users. $\hat{J}_i(\Thv_i)$ is obtained as $\hat{J}_i(\Thv_i)=\frac{1}{N_{r}}\sum_{i=1}^{N_{r}}\sum_{\x^{(0)}\in\X}\sum_{t=0}^{T}\beta^{t}c(\x^{(t)},\IND_{\{b_i^{(t)}\geq{\Th_{h_i}}\}})$ for user $i$. We take $\hat{J}(\Thv)=\sum_i\hat{J}_i(\Thv_i)$.} In this system, since $\Thv_i$ is a $8$-tuple for all $i\in\{1,\dotsc,5\}$, the dimension of $\Thv$ is $40$. We show the convergence performance of DSPSA, $\LN$-convex SA and CSPSA in Fig.~\ref{fig:CongestGBR}. The step size parameter $A$ is $0.63$ for DSPSA, $0.68$ for $\LN$-convex SA and $0.6$ for CSPSA. It can be seen that $\LN$-convex SA still has the best converge performance. But, unlike in the single-user system, DSPSA converges faster than CSPSA.

\subsubsection{Four-user Two Way Relay Channel}
\label{sec:multiuser2}

\begin{figure}[tpb]
	\centering
		\scalebox{0.9}{\begin{tikzpicture}

\draw [densely dotted,color=gray,fill=gray!20] (1.2,2.2) rectangle (4.8,-2.5);

\draw (2,2) rectangle (3.6,1.6);
\node at (2.8,1.8) {scheduler};
\draw [red,dotted,->] (2,1.8) --(1.4,1.8) -- (1.4,0.4) -- (1.7,0.4) -- (1.7,0.1);
\draw [red,dotted,->] (1.4,0.4) -- (1.4,-1.5) --  (1.7,-1.5) -- (1.7,-1.8);

\draw [red,dotted,->] (3.6,1.8) -- (4.7,1.8) -- (4.7,1.4) -- (4.3,1.4) -- (4.3,1);
\draw [red,dotted,->]  (4.7,1.4) -- (4.7,-0.6) -- (4.3,-0.6) -- (4.3,-0.9);


\draw (-1.2,0.7) rectangle (-0.6,0.1);
\node at (-0.9,0.4) {$1$};
\draw (-0.6,0.4) -- (1.2,0.4);

\draw (6.6,0.7) rectangle (7.2,0.1);
\node at (6.9,0.4) {$2$};
\draw (4.8,0.4) -- (6.6,0.4);

\draw [fill=red!20] (2.8,1.2) rectangle (3.4,0.5);
\draw (2.2,1.2) rectangle (3.4,0.5);
\draw (2.2,1.2) -- (1.8,1.2);
\draw (2.2,0.5) -- (1.8,0.5);
\draw (2.5,1.2) -- (2.5,0.5);
\draw (2.8,1.2) -- (2.8,0.5);
\draw (3.1,1.2) -- (3.1,0.5);
\draw [ ->] (1,0.85) -- (1.9,0.85);
\draw (3.4,0.85) -- (4.1,0.85);
\draw (4.1,0.85) -- (4.5,1.05);
\draw [->] (4.5,0.85) -- (5,0.85);
\node at (0.6,0.95) {\small $f_1^{(t)}$};

\draw [fill=blue!20](2.6,0.3) rectangle (3.2,-0.4);
\draw (3.8,-0.4) rectangle (2.6,0.3);
\draw (3.8,-0.4) -- (4.2,-0.4);
\draw (3.8,0.3) -- (4.2,0.3);
\draw (3.5,-0.4) -- (3.5,0.3);
\draw (3.2,-0.4) -- (3.2,0.3);
\draw (2.9,-0.4) -- (2.9,0.3);
\draw [<-] (4.1,-0.05) -- (5,-0.05);
\draw (2.6,-0.05) -- (1.9,-0.05);
\draw (1.9,-0.05) -- (1.5,0.15);
\draw [<-] (1,-0.05) -- (1.5,-0.05);
\node at (5.4,-0.15) {\small $f_2^{(t)}$};

\node at (0.4,0.55) {\scriptsize fading};
\node at (0.4,0.25) {\scriptsize channel $1$};
\node at (5.65,0.55) {\scriptsize fading};
\node at (5.65,0.25) {\scriptsize channel $2$};


\draw (-1.2,-1.2) rectangle (-0.6,-1.8);
\node at (-0.9,-1.5) {$3$};
\draw (-0.6,-1.5) -- (1.2,-1.5);

\draw (6.6,-1.2) rectangle (7.2,-1.8);
\node at (6.9,-1.5) {$4$};
\draw (4.8,-1.5) -- (6.6,-1.5);

\draw [fill=green!20] (2.8,-0.7) rectangle (3.4,-1.4);
\draw (2.2,-0.7) rectangle (3.4,-1.4);
\draw (2.2,-0.7) -- (1.8,-0.7);
\draw (2.2,-1.4) -- (1.8,-1.4);
\draw (2.5,-0.7) -- (2.5,-1.4);
\draw (2.8,-0.7) -- (2.8,-1.4);
\draw (3.1,-0.7) -- (3.1,-1.4);
\draw [ ->] (1,-1.05) -- (1.9,-1.05);
\draw (3.4,-1.05) -- (4.1,-1.05);
\draw (4.1,-1.05) -- (4.5,-0.85);
\draw [->] (4.5,-1.05) -- (5,-1.05);
\node at (0.6,-0.95) {\small $f_3^{(t)}$};

\draw [fill=orange!30](2.6,-1.6) rectangle (3.2,-2.3);
\draw (3.8,-2.3) rectangle (2.6,-1.6);
\draw (3.8,-2.3) -- (4.2,-2.3);
\draw (3.8,-1.6) -- (4.2,-1.6);
\draw (3.5,-2.3) -- (3.5,-1.6);
\draw (3.2,-2.3) -- (3.2,-1.6);
\draw (2.9,-2.3) -- (2.9,-1.6);
\draw [<-] (4.1,-1.95) -- (5,-1.95);
\draw (2.6,-1.95) -- (1.9,-1.95);
\draw (1.9,-1.95) -- (1.5,-1.75);
\draw [<-] (1,-1.95) -- (1.5,-1.95);
\node at (5.4,-2.05) {\small $f_4^{(t)}$};

\node at (0.4,-1.35) {\scriptsize fading};
\node at (0.4,-1.65) {\scriptsize channel $3$};
\node at (5.65,-1.35) {\scriptsize fading};
\node at (5.65,-1.65) {\scriptsize channel $4$};

\end{tikzpicture}}
	\caption{Four-user on-off transmission control problem in NC-TWRC \cite{Ding2012}. User 1 communicates with user 2, and user 3 communicates with user 4. A scheduler controls the outflows of all queues.}
	\label{fig:TWRC}
\end{figure}
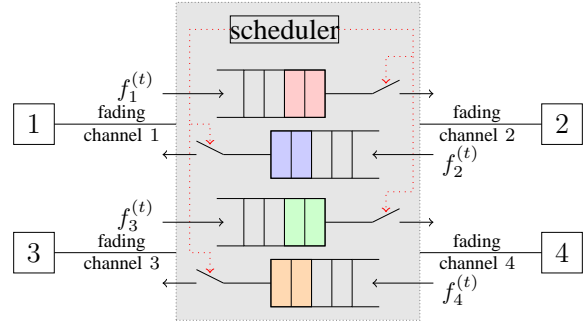

Fig.~\ref{fig:TWRC} shows a transmission control problem in a four-user NC-TWRC system. There are two pairs of users communicating with each other via the relay: User 1 exchanges packets with user 2; User 3 exchanges packets with user 4. A scheduler at the relay controls the downlink packet flows for four users. Network coding (XORing) is allowed in this system. If the scheduler decides to transmit one packet from each user in any pair, the two packets will be XORed and broadcast. Otherwise, the departing packets are simply forwarded to the destination. Take users 1 and 2 for example. If $a_1=a_2=1$, the departing packets from users $1$ and $2$ are XORed and broadcast in order to save transmission power; If $a_1=1,a_2=0$ or $a_1=0,a_2=1$, the relay simply forwards the packet to the destination. The same applies to users $3$ and $4$. We assume that the downlink channels are orthogonal so that the relay can simultaneously exchange packets for both pairs of users. We set $p_{f_i}=0.5$ for all $i\in\{1,\dotsc,4\}$.

In this system, the transmission control problem of each pair of users can be modeled by an MDP. We show the MDP model for users $1$ and $2$ as follows. The MDP for users $3$ and $4$ can be derived in the same way. The MDP model for users $1$ and $2$ has the system state $\x=(b_1,h_1,b_2,h_2)$ and action $\act=(a_1,a_2)\in\{0,1\}^2$. The state transition probability is $P_{\x\x'}^{\act}=\Pi_{i=1}^2P_{b_ib'_i}^{a_i}P_{h_ih'_i}$. The immediate costs is defined as
    \begin{equation}
        c(\x,\act)=\sum_{i=1}^{2} \Big( c_q(b_i,a_i) + c_{tr}(h_i,a_i) \Big) + \IND_{\{a_1=1 \text{ or }a_2=1\}}. \nonumber
    \end{equation}
Here, $c(\x,\act)$ contains two parts: the sum of $c_q$ and $c_{tr}$ incurred at both user $1$ and user $2$ and $\IND_{\{a_1=1 \text{ or }a_2=1\}}$ which is the cost that is proportional to the power consumption at the relay\cite{Ding2012}. In this MDP, the optimal policy contains two parts: $\theta_1^*$ and $\theta_2^*$. $\theta_1^*(\x)$ and $\theta_2^*(\x)$ determine the optimal action to queue $1$ and queue $2$, respectively, for a certain state $\x$. By following the same approach as the proof in Proposition~\ref{prop1}, one can show that function $Q(\x,\act)$ is submodular in $(b_i,a_i)$ for all $i\in\{1,2\}$. The optimal policies $\theta_1^*$ and $\theta_2^*$ are nondecreasing in $b_1$ and $b_2$, respectively. Likewise, for the MDP model for users $3$ and $4$, the optimal policy $\theta_3^*$ and $\theta_4^*$ is nondecreasing in $b_3$ and $b_4$, respectively. The optimal policy in the entire system is $\theta^*=(\theta_1^*,\dotsc,\theta_4^*)$.

\begin{figure}[tb]
	\centering
        \subfigure[{the value of the objective function $J([\tilde{\Thv}^{(n)}])$ vs. iteration}]{\scalebox{0.85}{\input{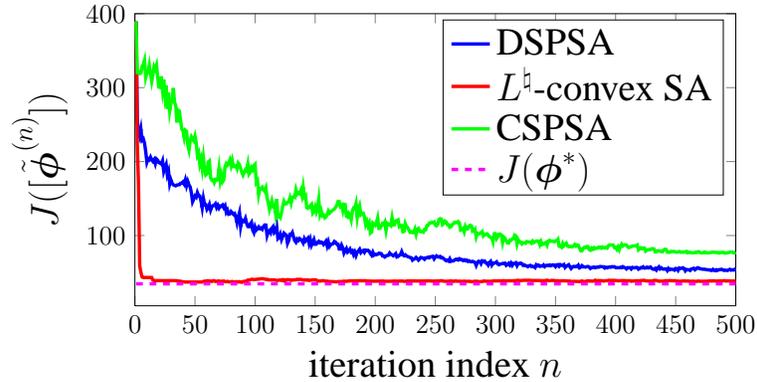}}} \quad
        \subfigure[the normalized error $\frac{\|\tilde{\Thv}^{(n)}-\Thv^*\|}{\|\tilde{\Thv}^{(0)}-\Thv^*\|}$ vs. iteration]{\scalebox{0.85}{\input{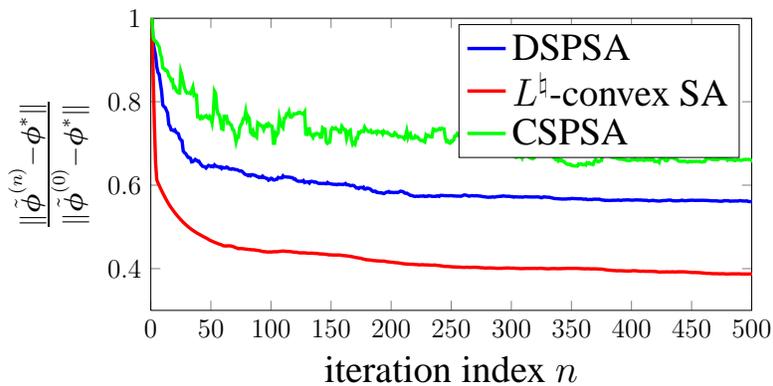}}}
	\caption{Convergence performances of DSPSA, $\LN$-convex SA and CSPSA in a four-user NC-TWRC system in Fig.~\ref{fig:TWRC}. The channels are modeled by $4$-state FSMC. In this system, $p_{f_i}=0.5$ and $L_i=5$ for all $i\in\{1,\dotsc,4\}$. The dimension of the threshold vector $\Thv$ is $320$. }
	\label{fig:QueCSI1}
\end{figure}

Let $\Thv_i$ be the queue threshold vector to queue $i$. $\Thv_1$ is constructed by stacking $\Th_{h_1b_2h_2}=\min\{b_1\colon\theta_1(\x)=1\}$ for all values of $(h_1,b_2,h_2)$, and $\Thv_2$ is constructed by stacking $\Th_{b_1h_1h_2}=\min\{b_2\colon\theta_2(\x)=1\}$ for all $(b_1,h_1,h_2)$. $\Thv_3$ and $\Thv_4$ are constructed in the same way. In this system, $\Thv_i$ is an $80$-tuple variable for all $i\in\{1,\dotsc,4\}$. The queue threshold vector in the entire system is $\Thv=(\Thv_1,\dotsc,\Thv_4)$, the dimension of which is $320$. We show the convergence performance of DSPSA, $\LN$-convex SA and CSPSA in Fig.~\ref{fig:QueCSI1}. The step size parameter $A$ is $0.43$ for DSPSA, $0.6$ for $\LN$-convex SA and $0.4$ for CSPSA. The results are similar as in Fig.~\ref{fig:CongestGBR}: $\LN$-convex SA converges faster than DSPSA, and DSPSA converges faster than CSPSA.

\subsection{Accuracy and Complexity}
\label{sec:complex}
We compare the DP and three SA algorithms, DSPSA, $\LN$-convex SA and CSPSA as follows.

\subsubsection{SA vs. DP}
Based on \eqref{eq:V}, the complexity in each iteration of DP is $O(|\X|^2\cdot|\A|)$.\footnote{There are $|\X|$ minimization operations, each of which requires $|\A|$ calculations of $Q$, and each $Q$ value requires $|\X|$ multiplications over state $\x'$.} Let $\alpha$ be the complexity of obtaining the simulated value of $\sum_{i=1}^{N_{r}}\sum_{t=0}^{T}\beta^{t}c(\x^{(t)},\IND_{\{b^{(t)}\geq{\Th_h}\}})$ in \eqref{eq:SimJ} and $D$ be the dimension of the threshold vector $\Thv$. In each iteration, the complexity of both DSPSA and CSPSA is $O(|\X|\cdot\alpha)$, and the complexity of $\LN$-convex SA is $O(D\cdot|\X|\cdot\alpha)$. Here, the complexity $\alpha$ is incurred by simulation instead of calculation. Also, $D$ is smaller than $|\X|$. For example, in the single-user system in Section~\ref{sec:singleuser}, the number of states is $|\X|=88$ and the dimension of $\Thv$ is $D=8$. Therefore, the complexity of SA algorithms is lower than that of DP.

Moreover, the three SA algorithms are simulation-based algorithms, the runs of which do not require the full knowledge of the MDP model. Based on \eqref{eq:SimJ}, to obtain $\hat{J}$, one only requires the knowledge of the state space and a simulation model that can generate a state sequence based on a given threshold policy and the statistics of packet arrival and channel variation processes. By SA algorithms, it is possible for the scheduler to learn the optimal policy online. For example, assume that we apply DSPSA to the single-user system in Section~\ref{sec:singleuser}. At the beginning, $\tilde{\Thv}^{(0)}$ is any arbitrary threshold policy. The scheduler adopts policies $\lfloor\tilde{\Thv}^{(0)}\rfloor+\frac{\One+\mathbf{\Delta}}{2}$ and $\lfloor\tilde{\Thv}^{(0)}\rfloor+\frac{\One-\mathbf{\Delta}}{2}$ for a while and obtains corresponding values of $\hat{J}$ based on the actual immediate costs incurred. It then obtains $\gv$ as in \eqref{eq:DSPSA} and adapts to the new threshold policy $\Thv^{(1)}$. By repeating this process, the scheduler can slowly update the policy towards the optimal one.

\subsubsection{DSPSA and CSPSA vs. $\LN$-convex SA}

From Figs.~\ref{fig:AMC2}, \ref{fig:CongestGBR} and \ref{fig:QueCSI1}, it can be seen that $\LN$-convex SA always converges faster than DSPSA and CSPSA. However, the complexity of $\LN$-convex SA depends on $D$, the dimension of $\Thv$, and can be much higher than DSPSA and CSPSA in multi-user systems. In each iteration, $\LN$-convex SA requires $D+1$ measurements of $\hat{J}$. For example, let $m$ be the total number of users in the system, and let $|\B|$ and $|\HSet|$ be the cardinality of the queue and channel states, respectively, that is associated with one user. For the single-user system in Section\ref{sec:singleuser}, $\LN$-convex SA requires $|\HSet|+1$ values of $\hat{J}$ in each iteration. But, $\LN$-convex SA requires $m\cdot|\HSet|+1$ and $\frac{m}{2}\cdot|\B|\cdot|\HSet|^2+1$ values of $\hat{J}$ in each iteration for the multi-user systems in Sections~\ref{sec:multiuser1} and \ref{sec:multiuser2}, respectively. If an $m$-user system is modeled by one MDP, e.g., the MDP model in \cite{Huang2010}, the dimension of $\Thv$ is $m|B|^{m-1}|\HSet|^m$, i.e., the complexity of $\LN$-convex SA may grow exponentially with the number of users. On the contrary, both DSPSA and CSPSA are perturbation-based algorithms which always require only $2$ measurements of $\hat{J}$ in each iteration for all systems no matter how large the state space of the MDP is. Therefore, the complexity of DSPSA and CSPSA is much lower than $\LN$-convex SA in multi-user systems.

\subsubsection{DSPSA vs. CSPSA}

DSPSA is the algorithm that is directly proposed for discrete convex minimization problems. The gradient $\gv$ in DSPSA is obtained based on the definition of mid-point convexity\cite{Wang2011}. CSPSA is a discrete version of an SA algorithm that is originally proposed for continuous minimization problems. 
Based on Fig.~\ref{fig:AMC2}, DSPSA converges at the same speed as CSPSA in the single-user system. But, based on Figs.~\ref{fig:CongestGBR} and \ref{fig:QueCSI1}, the convergence performance of DSPSA is better than that of CSPSA in multi-user systems. However, even if the performance of DSPSA is comparable to CSPSA in the single-user system, it should be noted that DSPSA is simpler to implement than CSPSA. In CSPSA, a projection function $\Gamma$ in \eqref{eq:project} is used. The idea of $\Gamma(\tilde{\Thv})$ is to treat the real-valued $\tilde{\Thv}$ as a threshold policy that is a randomized mixture of two deterministic ones, $\lfloor{\tilde{\Thv}}\rfloor$ and $\lceil{\tilde{\Thv}}\rceil$. On the contrary, in DSPSA, the scheduler only needs to follow one deterministic policy to for each value of $\hat{J}$. Therefore, although both DSPSA and CSPSA require $2$ measurements of $\hat{J}$ in each iteration, the complexity of DSPSA is lower than that of CSPSA since DSPSA does not need to implement the projection function.

The results above can be used to guide the implementation of the SA algorithms in practical systems. For example, if one finds that the discrete convexity exists in some cross-layer on-off transmission control system, an SA algorithm with lower complexity than DP may be run to approximate the optimal policy. Also, it is better to choose DSA algorithms that is directly proposed for the discrete convexity, e.g., DSPSA and $\LN$-convex SA, rather than CSPSA. In a multi-user system, since the complexity of $\LN$-convex SA is high, one can just implement DSPSA which achieves best trade-off between accuracy and complexity. It should also be pointed out that the main contribution of this paper is the formulation and proof of discrete convexity of the minimization problem~\eqref{eq:LConvexOpt}. The solution of this problem is not restricted to the DSPSA, $\LN$-convex SA or CSPSA presented in this paper. One may find more efficient algorithms in discrete stochastic minimization literature, and the results derived in this paper can be used to show the global and almost sure convergence. For example, if an algorithm has higher accuracy than DSPSA and lower complexity than $\LN$-convex SA but only converges to local optimizer, then one directly know it converges globally when it applies to problem~\eqref{eq:LConvexOpt}.

\section{Conclusion}
In this paper, we formulated a multivariate minimization problem for searching the optimal queue threshold policy in a cross-layer on-off transmission control system. We proved that the objective function of minimization problem was both discrete separable convex and $\LN$-convex if the DP was submodular. We proposed to use two DSA algorithms, DSPSA and $\LN$-convex SA, to approximate the optimal policy. We applied the two DSA algorithms and a CSPSA algorithm in single-user and multi-user systems. The results showed that: $\LN$-convex SA always converged faster than DSPSA and CSPSA;  DSPSA converged faster than CSPSA in multi-user systems. We also analyzed the complexity of the two DSA and CSPSA algorithms to show that: the complexity of $\LN$-convex SA grew much higher than DSPSA and CSPSA in multi-user systems; The complexity of DSPSA was lower than CSPSA.

Finally, we point out two possible extensions of the work in this paper: One may design more efficient SA or stochastic optimization
algorithms based on the discrete convexity of the on-off transmission control problem, e.g., an SA algorithm that is more accurate than DSPSA and involves less complexity than $\LN$-convex SA; It would be of interest if the convexity of the optimization problem can be found in a variable rate cross-layer adaptive modulation system, e.g., cross-layer $m$-QAM modulation system.

\appendices

\section{}
\label{appAMC1}

Assume that $V(b',h')$ is nondecreasing and convex in $b'$. Define
\begin{equation}
    \varphi(y,f,h') = w\Big[ [y]^{+}+f-L \Big]^{+}   + \beta V(\min\{[y]^{+}+f,L\},h'). \nonumber
\end{equation}
Then, $Q(b,h,a)=c_{tr}(h,a)+ \sum_{h'}P_{hh'} \E_f [ \varphi(b-a,f,h') ]$. Consider the convexity of $\varphi$ in $y$. We have
\begin{align}
    & \quad \varphi(y+1,f,h')+\varphi(y-1,f,h')-2\varphi(y,f,h')    \nonumber \\
    &=\begin{cases}
            0 & y=-1\\
            \beta(V(1+f,h') -V(f,h')) \geq{0} & y=0  \\
            w+\beta (V(L-1,h')-V(L,h') & y+f=L \\
            0 & y+f=L+1 \\
            \beta \Big( V(y+1+f,h') +V(y-1+f,h') -V(y+f,h') \Big) \geq{0} & \text{otherwise}
        \end{cases}.  \nonumber
\end{align}
Let $a^*(b,h)=\arg\min_a Q(b,h,a)$. Then, $V(b,h)=Q(b,h,a^*(b,h))$. Since
\begin{align}
    &\quad w+\beta(V(L-1,h')-V(L,h'))   \nonumber \\
    &=w+\beta(Q(L-1,h',a^*(L-1,h')) -Q(L,h',a^*(L,h')))   \nonumber \\
    &\geq w+\beta(Q(L-1,h',a^*(L-1,h'))  -Q(L,h',a^*(L-1,h'))) \nonumber \\
    &\geq w(1-\beta) \geq 0,
\end{align}
$\varphi$ is convex in $y$. Consider the submodularity of $Q$ in $(b,a)$. Since
\begin{align}
    &\quad Q(b+1,h,0) + Q(b,h,1) - Q(b,h,0) - Q(b,h,1)  \nonumber \\
        &=\sum_{h'}P_{hh'} \E_f \Big[ \varphi(b+1,f,h') + \varphi(b,f,h') - 2\varphi(b-1,f,h') \Big] \geq{0},
\end{align}
based on Definition~\ref{def:submodular}, $Q$ is submodular in $(b,a)$ for all $h$. Consider the monotonicity of $\varphi$ in $y$. It is straightforward to see that both $[[y]^{+}+f-L]$ and $\min\{[y]^{+}+f,L\}$ are nondecreasing in $y$ for all $(f,h')$. Since $V$ is nondecreasing in $b'$, $\varphi$ is nondecreasing in $y$. We have
\begin{align}
    &\quad Q(b+1,h,a)-Q(b,h,a)    \nonumber \\
    & = \sum_{h'}P_{hh'} \E_f \Big[ \varphi(b-a+1,f,h') - \varphi(b-a,f,h') \Big] \geq{0}.   \nonumber
\end{align}
Therefore, $Q$ is nondecreasing in $b$ for all $(h,a)$. \hfill \IEEEQED

\section{}
\label{appAMC2}
Assume that $Q$ is submodular in $(b,a)$ and nondecreasing in $b$. Due to the submodularity of $Q$ in $(b,a)$,
\begin{equation}
    a^*(b+1,h) \geq a^*(b,h) \geq a^*(b-1,h).   \nonumber
\end{equation}
It is easy to see that $c_{tr}$ defined in \eqref{eq:Ctr1} is convex in $a$ for all $h$. Recall that the submodularity of $Q$ in $(b,a)$ is equivalent to the convexity of $\varphi$ in $y$. We have
\begin{align} \label{eq:app1}
    &\quad V(b+1,h) + V(b-1,h) - 2V(b,h) \nonumber \\
    &=Q(b+1,h,a^*(b+1,h))+Q(b-1,h,a^*(b-1,h))  -2Q(b,h,a^*(b,h))   \nonumber \\
    &\geq Q(b+1,h,a^*(b+1,h))+Q(b-1,h,a^*(b-1,h))   \nonumber \\
    &\quad -Q(b,h,a^*(b+1,h))-Q(b,h,a^*(b-1,h)).
\end{align}
For $a^*(b+1,h)=a^*(b-1,h)+1$, \eqref{eq:app1} equals $0$, and for $a^*(b+1,h)=a^*(b-1,h)$, \eqref{eq:app1} is greater or equal to $0$.
Therefore, $V$ is convex in $b$. Recall that the monotonicity of $Q$ in $b$ is given by the monotonicity of $\varphi$ in $y$. We have
\begin{align}
    &\quad V(b+1,h) - V(b,h) \nonumber \\
    &=Q(b+1,h,a^*(b+1,h))-Q(b,h,a^*(b,h))   \nonumber \\
    &\geq Q(b+1,h,a^*(b+1,h))-Q(b,h,a^*(b+1,h))  \nonumber \\
    &= \sum_{h'}P_{hh'} \E_f \Big[ \varphi(b+1-a^*(b+1,h),f,h') - \varphi(b-a^*(b+1,h),f,h') \Big] \geq{0}.
\end{align}
Therefore, $V$ is nondecreasing in $b$ for all $h$.   \hfill \IEEEQED

\bibliography{IEEEabrv,DSAbib}

\end{document}